\documentclass[nologo,11pt,a4paper]{ETHpaper}
\usepackage[numbers,compress, sort]{natbib}
\usepackage{setspace}
\usepackage{graphicx, amsmath, amssymb,color,wasysym}
\usepackage{mathtools}
\usepackage{nicefrac}
\usepackage{float}
\usepackage{booktabs}
\usepackage{adjustbox}
\usepackage{array}
\usepackage{etoolbox}
\usepackage{pdflscape}
\usepackage{lineno,hyperref}
\usepackage{graphicx}
\usepackage{changepage}
\usepackage{tabularx,booktabs}
\usepackage{lscape}
\usepackage{rotating}
\usepackage{makecell}
\usepackage{subfig}
\makeatletter
\let\@fnsymbol\@arabic
\makeatother
\usepackage[bottom,flushmargin,hang,multiple]{footmisc}
\usepackage{etoolbox}
\makeatletter
\def\blfootnote{\xdef\@thefnmark{}\@footnotetext}
\makeatother
\let\oldmaketitle\maketitle
\renewcommand{\maketitle}{\oldmaketitle\setcounter{footnote}{0}}
\usepackage{verbatim}

\begin{document}

\title{The development of nations conditions the disease space}
\renewcommand{\thefootnote}{\fnsymbol{footnote}}
\titlealternative{The development of nations conditions the disease space}

\author{Antonios Garas$^{1}$, Sophie Guthmuller$^{2}$, \\ and Athanasios Lapatinas$^{2}$\footnote{Corresponding author (\href{mailto: athanasios. lapatinas@ec.europa.eu}{athanasios.lapatinas@ec.europa.eu}). This research was conducted while Sophie Guthmuller and Athanasios Lapatinas were in service at the European Commission's Joint Research Centre. The scientific output expressed does not imply a European Commission policy position. Neither the European Commission nor any person acting on behalf of the Commission is responsible for any use that might be made of this publication.}}
\authoralternative{A. Garas, S. Guthmuller, and A. Lapatinas}
\address{
		$^1$ETH Zurich, Chair of Systems Design, \\Weinbergstrasse 56/58, 8092 Zurich, Switzerland\\
		$^2$European Commission, Joint Research Centre, \\Via E. Fermi 2749, TP 361, Ispra (VA), I-21027, Italy\\
  	}
\www{\url{http://www.sg.ethz.ch}}
\makeframing
\maketitle

\renewcommand{\thefootnote}{\arabic{footnote}}

\begin{abstract}
Using the economic complexity methodology on data for disease prevalence in 195 countries during the period of 1990-2016, we propose two new metrics for quantifying the relatedness between diseases, or the `disease space' of countries. With these metrics, we analyze the geography of diseases and empirically investigate the effect of economic development on the health complexity of countries. 
We show that a higher income per capita increases the complexity of countries' diseases. Furthermore, we build a disease-level index that links a disease to the average level of GDP per capita of the countries that have prevalent cases of the disease. With this index, we highlight the link between economic development and the complexity of diseases and illustrate, at the disease-level, how increases in income per capita are associated with more complex diseases.  

\mbox{}
\end{abstract}

\doublespacing

\clearpage
\newpage

\section{Introduction}
\label{sec: intro}

 Popular belief holds that the European conquest of America was accomplished with guns and soldiers. However, \citet{bianchine1992role} show that new illnesses brought from the Old World by European conquistadors, which resulted in devastating epidemics throughout the New World, were the major forces behind the aboriginal depopulation of the Americas. Our history, geography, culture, religion, and language have often been influenced by infections that have plagued humankind and shaped important events. Examples include the plague in fourteenth century Europe, how the yellow fever increased the importation of African slaves in the sixteenth century due to shortage of indigenous workers and the relative resistance of Africans to the disease, as well as the typhus deaths of the Napoleon's army during the 1812 Russian campaign, and Franklin D. Roosevelt's hypertension and heart failure, which worsened during his February 1945 dealings with Joseph Stalin in Malta \citep{post1995illness, sherman2007twelve, taylor2016white}.     

Furthermore, there is strong historical evidence that the wealth of nations is positively linked to the health of their populations. Since the eighteenth century, economic development associated with improvements in nutrition, access to sanitation, public health interventions, and medical innovations such as vaccination, have contributed to the reduction of major infectious diseases, the decline of premature death rates, and a longer life expectancy for children and adults in both developed and developing countries \citep{Fogel1990, fogel2004, BIRCHENALL2007}. 

Nevertheless, many significant health problems have emerged in concert with economic development and technological modernization. Among them, stress, anxiety, sleep deprivation, and depression are mental disorders that are more prevalent in high-income countries. While they account for only 9\% of the burden in low-income countries, this figure is 18\% in middle-income and 27\% in high-income countries \cite{prince2011}.  In OECD countries, a longer life expectancy is coupled with a higher rate of chronic and long-term illnesses in older populations \cite{GBD2016dementia}. Industrialization has expanded the reach of existing food-related diseases and created new disorders and addictions \cite{GBD2015obesity}. Industrialization also stimulates urbanization, the process of population migration from rural areas to cities. This makes urban areas focal points for many emerging environmental and health hazards. According to the World Health Organization (WHO), ``\textit{as urban populations grow, the quality of global and local ecosystems, and the urban environment, will play an increasingly important role in public health with respect to issues ranging from solid waste disposal, provision of safe water and sanitation, and injury prevention, to the interface between urban poverty, environment and health.}''.\footnote{WHO, `Urbanization and health' \url{https://www.who.int/globalchange/ecosystems/urbanization/en/}} Industrialization is also linked to occupational accidents and work-related diseases (e.g., work-related cancers, musculoskeletal disorders, respiratory diseases, psycho-social problems, and circulatory diseases), which are worldwide problems resulting in important losses for individuals, organizations and societies \citep{driscoll2005review, nelson2005global, steenland2003dying, park2002alternate, zahm2003occupational, hansen2001increased, hoy2014global, punnett2005estimating, hoy2010measuring}.     

From the above discussion, it becomes clear that economic development can affect population health in a number of ways, both positive and negative. To disentangle the net impact of economic development on countries' health status, we develop a new metric called the Health Complexity Index ({\textrm HCI}), which quantifies the disease space of countries, i.e., the network representation of the relatedness and proximity between diseases with prevalent cases worldwide. To compute the  {\textrm HCI}, we follow the economic complexity methodology, which was initially applied to trade micro-data, measuring the  amount of knowledge materialized in a country's productive structure. 

More, specifically, the Economic Complexity Index ({\textrm ECI}) is a metric that quantifies a country's product space, i.e., the network of products traded internationally. When a country produces a good that is located in the core of the product space, many other related goods can also be produced with the given capabilities. However, this does not hold for goods lying in the network's periphery, because they require different capabilities. The $ECI$ methodology encapsulates this information by assigning lower values to countries that export products located at the periphery of the product space and higher values to countries that export commodities located in the center of the product space \citep{hidalgo2007product}. 

Based on the {\textrm ECI} methodology, a number of recent contributions explain economic development and growth as a process of information development and of learning how to produce and export more diversified products \citep{abdon2011product,bustos2012dynamics,caldarelli2012network,
cristelli2013measuring,felipe2012inclusive,
hausmann2007you,hausmann2014atlas,hidalgo2009building,hidalgo2007product, 
rodrik2006s,tacchella2013economic,albeaik2017measuring, saviotti2008export,cristelli2015heterogeneous,hausmann2011network}. Furthermore, \citet{hartmann2017linking} have recently shown that countries exporting complex products tend to be more inclusive and have lower levels of income inequality than countries exporting simpler products. In addition, \citet{Lapatinas2018} find that countries with high intellectual quotient (IQ) populations produce and export more sophisticated/complex products, while \citet{lapatinas2019effect} shows that the Internet has a positive effect on economic complexity. Adopting the economic complexity methodology, \citet{balland2017} compute a knowledge complexity index with more than two million patent records for US metropolitan areas between 1975-2010. They analyze the geography and evolution of knowledge complexity in US cities and show that the most complex cities in terms of patents are not always those with the highest rates of patenting. In addition, using citation data, they show that more complex patents are less likely to be cited than simpler patents when the citing and cited patents are located in different metropolitan areas. 

In this paper, we build a complexity index that measures the composition of a country's pool of prevalent cases of diseases by combining information on the diversity of diseases in the country and the ubiquity of its diseases (the number of other countries that also have prevalent cases of that disease). The intuition is that relatively high scores on the health complexity index indicate populations that are diverse and have diseases that, on average, have low ubiquity, i.e., these diseases have prevalent cases in only a few other countries.

In this view, the health complexity index does not refer to a complex treatment or to complex causes of a disease, but measures instead whether a disease is located in the densely connected core of the disease space i.e., whether many other related diseases have prevalent cases in many other countries. The country-disease network and the disease space reveal information about the health-related habits of populations, such as, lifestyle and dietary habits. There are also multiple reasons to expect the disease structures to be associated with their `structural transformations' (i.e., the industrialization process by which economies diversify from agriculture to manufacturing and services \citep{hausmann2006structural, ngai2007structural, gollin2002role, herrendorf2014growth, laitner2000structural}), with their environmental performance \citep{kunzli2000public, holgate1999air, chay2003impact, kampa2008human}, or with their adopted health-related policies \citep{glasgow1999evaluating, whitlock2002evaluating, mckenzie2005planning, drummond2015methods, folland2007economics}, as these contribute to their health status and living standards \citep{mankiw1992contribution, friedman2006moral}.

The aim of this paper is fourfold: $(i)$ to build two new metrics that quantify the disease space, following the economic complexity methodology; $(ii)$ to estimate the effect of economic development on countries' health complexity using the new metrics and following dynamic panel data econometric techniques; $(iii)$ to develop a disease-level index that links a disease to the average level of $GDP$ $per$ $capita$ of the countries in which the disease has prevalent cases; $(iv)$ to illustrate how a country's economic development is associated with changes in its disease composition and verify the relationship between economic development and health complexity at the disease level.

The remainder of the paper is structured as follows. Section \ref{sec: network} describes the data on disease prevalence and the construction of the \emph{country-disease network} and the disease space which form the analytical backbone of our study. Section \ref{sec: method} presents the methodology for developing the Health Complexity Index  {\textrm (HCI)} and the Disease Complexity Index {\textrm (DCI)}. Section \ref{sec: geography} presents the results of the structural analysis of the disease space and the country-disease network, with a particular focus on countries and regions. Section \ref{sec: econometrics} empirically investigates the effect of economic development on health complexity using the {\textrm HCI}, data on $GDP$ $per$ $capita$ and potential covariates. Section \ref{sec: economic} introduces an index that decomposes economic performance at the disease level. Using this index, we highlight the link between disease complexity and economic development. We demonstrate, at the disease level, that better economic performance is associated with more complex diseases. Finally, in section \ref{sec: concl}, we offer some concluding remarks.

\section{The country-disease network}
\label{sec: network}

\subsection{Data on prevalent cases of diseases}
\label{sec: data}

Information on diseases comes from the 2016 Global Burden of Diseases (GBD) study by the Institute for Health Metrics and Evaluation (IHME), an independent population health research center at UW Medicine (University of Washington) \citep{GlobalBurdenofDiseaseCollaborativeNetwork2017} that collects data from various sources to examine, among other things, the prevalence of diseases and injuries across the world (\url{http://www.healthdata.org/}).

Diseases and injuries are grouped by causes. The broader classification of causes (level 1) includes:  (a) \textit{communicable, maternal, neonatal, and nutritional diseases} such as HIV/AIDS and sexually transmitted infections, respiratory infections and tuberculosis, enteric infections (e.g., diarrheal diseases, typhoid fever), neglected tropical diseases (e.g. malaria, chagas disease) and other infectious diseases (e.g. meningitis and acute hepatitis), maternal and neonatal disorders (e.g., maternal abortion and miscarriage, ectopic pregnancy, maternal obstructed labor and uterine rupture), nutritional deficiencies (e.g., protein-energy malnutrition, vitamin A, iron, iodine deficiencies); (b) \textit{non-communicable diseases} such as cancers, cardiovascular diseases, chronic respiratory diseases, digestive diseases (e.g., cirrhosis, gastritis, pancreatitis), neurological disorders (e.g., multiple sclerosis, epilepsy, Parkinson's and Alzheimer's diseases, migraine), mental disorders (e.g., schizophrenia, anorexia nervosa and bulimia nervosa, conduct and hyperactivity disorders), substance use disorders (e.g., alcohol and drug use disorders), diabetes, kidney diseases, skin diseases (e.g., dermatitis, bacterial skin diseases), sense organ diseases (e.g., glaucoma, cataract, vision loss), musculoskeletal disorders (e.g., osteoarthritis, rheumatoid arthritis); (c) \textit{injuries} such as transport injuries (e.g., pedestrian road injuries, cyclist and motorcyclist road injuries), unintentional injuries (e.g., falls, poisonings, exposure to mechanical forces), self-harm and interpersonal violence (e.g., sexual violence, conflict and terrorism, executions).\footnote{In the remainder of the paper we use the word `disease' to refer to all diseases and injuries classified in the GBD study.} 

We use information for the most detailed level of causes in the GBD taxonomy (i.e., level 4, and when there is no level 4 classification, we use level 3). For example, among the non-communicable diseases (level 1), neoplasms (level 2) include the following level 3 categories: lip and oral cavity cancer, nasopharynx cancer, other pharynx cancer, esophageal cancer, stomach cancer, colon and rectal cancer, liver cancer, gallbladder and biliary tract cancer, pancreatic cancer, larynx cancer, etc. Then, liver cancer includes the following level 4 subcategories: liver cancer due to hepatitis B, liver cancer due to hepatitis C, liver cancer due to alcohol use, liver cancer due to non-alcoholic steatohepatitis (NASH), liver cancer due to other causes. In this case, as level 4 categories are available, we consider the information for these categories.    
 
Two measures of disease prevalence are exploited: the rate of prevalence (number of cases per 100,000 population) for all ages, and the age-standardized rate of prevalence to account for the differences in age structures across countries. This is useful because relative over- or under-representation of different age groups can obscure comparisons of age-dependent diseases (e.g., ischemic heart disease or malaria) across populations.

\subsection{The country-disease bipartite network}

Instrumental to our analysis is the bipartite network mapping of countries and diseases.
Bipartite, or bi-modal networks are abundant in the scientific literature, with examples including the city-tech knowledge network~\cite{balland2017}, the city-firm network~\cite{Schweitzer2019}, firm-projects networks~\cite{Balland2012}, predator-prey networks~\cite{allesina2012stability}, plants-pollinator networks~\cite{bascompte2003nested} etc.
Here, we use data from the 2016 Global Burden of Diseases study that assessed the disease burden of countries in the period of 1990 to 2016, and we generate an $l \times k$ country-diseases matrix $\bf E$, were the matrix element $E_{cd}$ represents the prevalent cases for disease $d$ per 100,000 population in country $c$.

The aforementioned matrix allows for the construction of an undirected, weighted county-disease network by linking each disease to the countries that have prevalent cases. 
These networks are very dense, and in order to visually explore their structure, we apply the Dijkstra algorithm~\cite{dijkstra1959note} to extract a Maximum Spanning Tree (MST) that  summarizes their structures.
More precisely, the MST, which is usually considered as the backbone of the network, is a connected subgraph having $l + k - 1$ edges with the maximum total weight and without forming any loops. 

\begin{figure}[t]
\centering
\includegraphics[width = 0.99\textwidth]{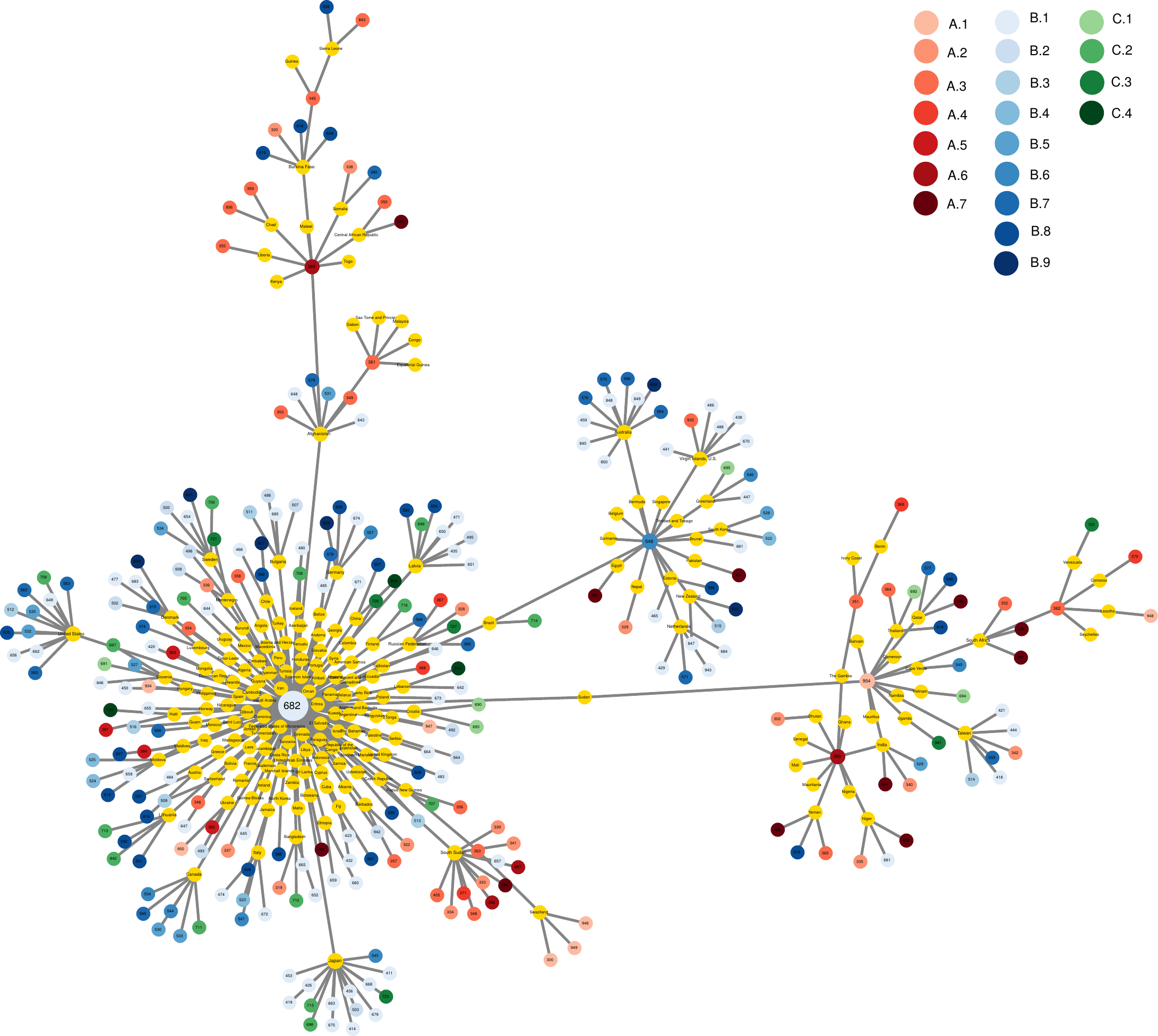}
\caption{{\bf Maximum Spanning Tree of the country-disease bipartite network.} 
\footnotesize Countries are represented by yellow nodes, and diseases cover the following categories: [A. `Communicable, maternal, neonatal, and nutritional diseases'] A.1 `HIV/AIDS and tuberculosis', A.2 `Diarrhea, lower respiratory, and other common infectious diseases', A.3 `Neglected tropical diseases and malaria', A.4 `Maternal disorders', A.5 `Neonatal disorders', A.6 `Nutritional deficiencies', A.7 `Other communicable, maternal, neonatal, and nutritional diseases'; [B. `Non-communicable diseases'] B.1 `Neoplasms', B.2 `Cardiovascular diseases', B.3 `Chronic respiratory diseases', B.4 `Cirrhosis and other chronic liver diseases', B.5 `Digestive diseases', B.6 `Neurological disorders', B.7 `Mental and substance use disorders', B.8 `Diabetes, urogenital, blood, and endocrine diseases', B.9 `Musculoskeletal disorders'; [C. `Injuries'] C.1 `Transport injuries', C.2 `Unintentional injuries', C.3 `Self-harm and interpersonal violence', C.4 `Forces of nature, conflict and terrorism, and executions and police conflict'. Data for 2016.}
\label{fig:CDnetwork}
\end{figure}

In Figure \ref{fig:CDnetwork} we illustrate the country-disease MST for 2016.
From this MST, we can easily identify clusters of countries that are linked to specific types of diseases. The main node of the network is caries in permanent teeth (disease cause number 682). 
This disease is the most common disease across the world, as it is present in the majority of countries. 
It is also the disease with the highest number of prevalent cases worldwide (2.44 billion cases in 2016 \cite{GlobalBurdenofDiseaseCollaborativeNetwork2017})

\subsection{The disease space}

The clustering of countries and diseases in the MST of the country-disease network already points towards relations in the prevalence of different diseases.
To explore this further, we construct the disease space, similar to the product-space introduced by~\citet{hidalgo2007product}.
More precisely, from the  country-disease matrix $\bf E$, we calculate the `{\it relative disease disadvantage}' (RDD) matrix, as described in the methods section (Section \ref{sec: method}).
In total, a country $c$ has a relative disease disadvantage in a particular disease $d$ if the proportion of prevalent cases of disease $d$ in the country's total pool of prevalent disease cases is higher than the proportion of prevalent cases of disease $d$ in the pool of prevalent disease cases in the rest of the world.
In this case, $\mathrm{RDD_{cd}}\geq 1$.

\begin{figure}[t]
\includegraphics[width = 1.0\textwidth]{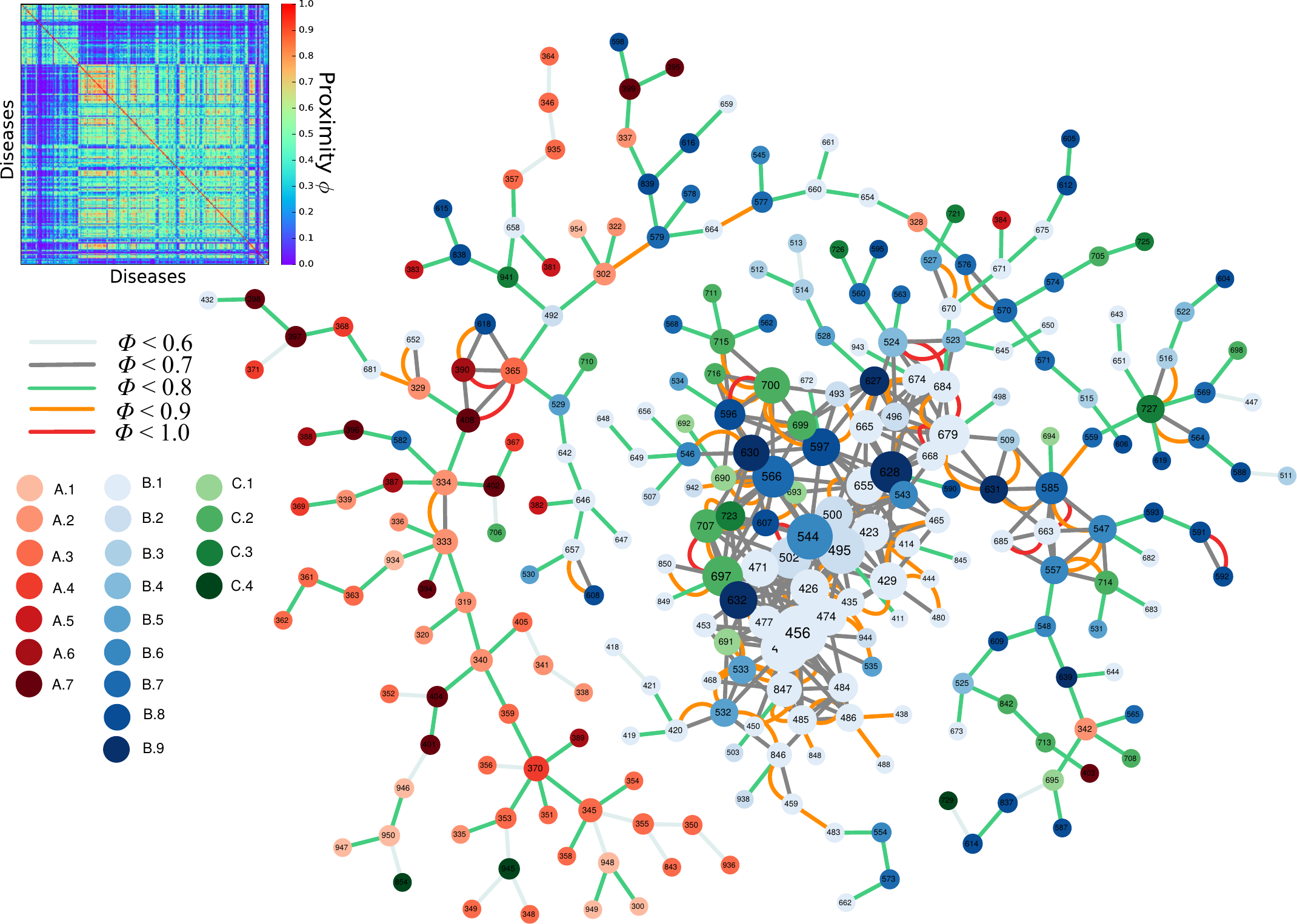}
\caption{{\bf The proximity matrix and the resulting disease space.} 
\footnotesize The size of the nodes is proportional to their degree, i.e., the number of links. Disease colors cover the following categories: [A. `Communicable, maternal, neonatal, and nutritional diseases'] A.1 `HIV/AIDS and tuberculosis', A.2 `Diarrhea, lower respiratory, and other common infectious diseases', A.3 `Neglected tropical diseases and malaria', A.4 `Maternal disorders', A.5 `Neonatal disorders', A.6 `Nutritional deficiencies', A.7 `Other communicable, maternal, neonatal, and nutritional diseases'; [B. `Non-communicable diseases'] B.1 `Neoplasms', B.2 `Cardiovascular diseases', B.3 `Chronic respiratory diseases', B.4 `Cirrhosis and other chronic liver diseases', B.5 `Digestive diseases', B.6 `Neurological disorders', B.7 `Mental and substance use disorders', B.8 `Diabetes, urogenital, blood, and endocrine diseases', B.9 `Musculoskeletal disorders'; [C. `Injuries'] C.1 `Transport injuries', C.2 `Unintentional injuries', C.3 `Self-harm and interpersonal violence', C.4 `Forces of nature, conflict and terrorism, and executions and police conflict'. Data for 2016.}
\label{fig:network}
\end{figure}

Calculating the RDD for all country-disease pairs allows us to derive a  matrix $\bf \Phi$, whose  elements $\Phi_{i,j}$ define a proximity measure between all pairs of diseases. 
This proximity measure reveals diseases that are prevalent in tandem, or in other words, with $\bf \Phi$, we measure the probability that a country $c$, which has a relative disease disadvantage in disease $i$, also has a relative disease disadvantage in disease $j$. The proximity measure is defined as:
\begin{equation}
    \Phi_{i,j}=\mathrm{min}\{ \mathrm{Pr}\mathrm{(RDD}_{i} \geq 1 \  | \  \mathrm{RDD}_{j} \geq 1), \ \mathrm{Pr}(\mathrm{RDD}_{j} \geq 1 \  | \  \mathrm{RDD}_{i} \geq 1) \},
\end{equation}
where $\mathrm{Pr}(\mathrm{RDD}_{i} \geq 1 \ | \  \mathrm{RDD}_{j} \geq 1)$ is the conditional probability of having a relative disease disadvantage in disease $i$ if you have a relative disease disadvantage in disease $j$.
Using the minimum of both conditional probabilities, we avoid issues of a rare disease having prevalent cases in only one country. Additionally, we make the resulting matrix $\bf \Phi$ symmetric (see Figure \ref{fig:network}). The proximity matrix is highly modular and its block structure reveals the presence of `communities', i.e., groups of diseases that are expected to occur together.

Next, we map this matrix onto a network, where each disease is represented by a node and every matrix element represents a weighted and undirected link.
Similar to the previous section, we start by applying Dijkstra's algorithm on matrix $\bf \Phi$ which calculates the MST of the network.
Following the rationale of~\citet{hidalgo2007product}, we start from the strongest links that are not part of the MST and keep adding links to the network until the average degree is four.
The resulting network is a visual representation of the disease space, which is shown in Figure~\ref{fig:network}.

\begin{figure}[t]
\centering
\includegraphics[width = 0.72\textwidth]{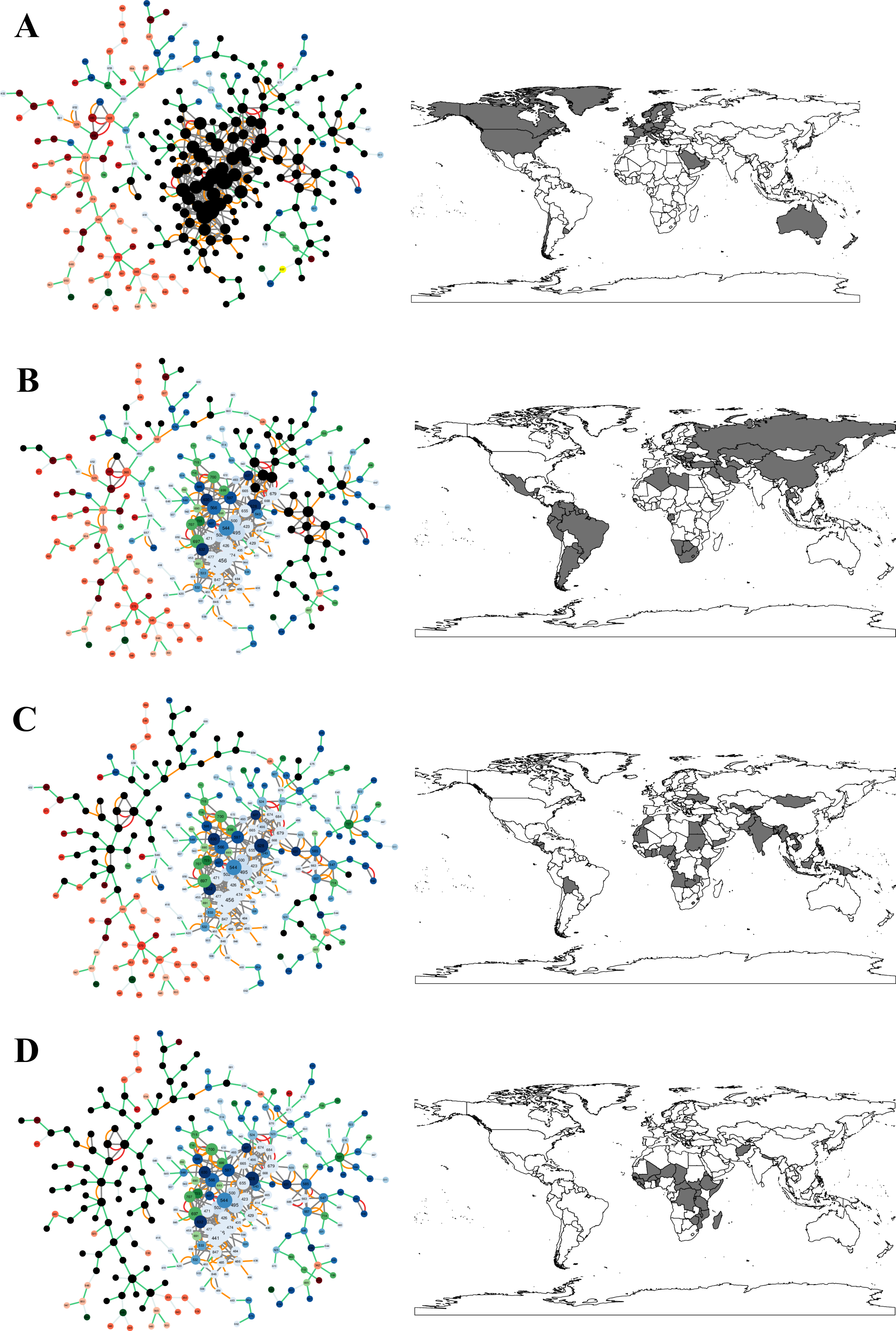}
\caption{{\bf Localization of diseases for different income-regions of the world.} 
\footnotesize {A: High income, B: Upper-middle income, C: Lower-middle income, D: Low income. Diseases in an income-region where more than half of the countries belonging to this region have a RDD>1 are shown with black nodes in the disease space network. Data for 2016.}}
\label{fig:network2}
\end{figure}

From Figure~\ref{fig:network}, it is evident that in the disease space network, different disease categories are clustered together and, similar to the product space network of~\citet{hidalgo2007product}, the network is heterogeneous and follows a core-periphery structure.
The external part of the network (the periphery) is mostly dominated by `communicable, maternal, neonatal, and nutritional diseases'. In Section \ref{sec: geography}, we show that these diseases are mostly prevalent in low-income countries. On the other hand, the core of the network is dominated by `non-communicable diseases', which have more prevalent cases in high-income countries (see Figure~\ref{fig:network2}).

\section{Methods}
\label{sec: method}
\subsection{Health complexity index}
To calculate \emph{health complexity} and \emph{disease complexity}, we combine information on prevalent cases of diseases and how common these diseases are across countries, following the economic complexity methodology, i.e., the formulas in the pioneering work of \citet{hidalgo2009building}.
In short, let us assume that we have disease information for $l$ number of countries and $k$ diseases.
With this information, we can fill an $l \times k$ diseases matrix {\bf E}, so that matrix element $E_{cd}$ is country $c$'s information for disease $d$.
If there is no information for disease $d$ in country $c$, then $E_{cd}=0$.
From this matrix, it is easy to calculate the following ratio:
\begin{equation}
 \mathrm{RDD}_{cd}= \frac{\nicefrac{X_{cd}}{\sum_{d'}X_{cd'}}}{\nicefrac{\sum_{c'}X_{c'd}}{\sum_{c'd'}X_{c'd'}}},
 \label{eq: rda}
\end{equation}
where $X_{cd}$ is the number of prevalent cases of disease $d$ per 100,000 population in country $c$.

Similar to the economic complexity methodology and the discussion in \citep{hidalgo2009building,caldarelli2012network,hartmann2017linking}, we claim that a country has a relative disease disadvantage in a disease when $\mathrm{RDD_{cd}}\geq 1$. In other words, a country $c$ has a {\textrm RDD} in disease $d$ if the proportion of prevalent cases of disease $d$ in the country's pool of all prevalent cases of disease is higher than the proportion of prevalent cases of disease $d$ in the world's pool of all prevalent cases of disease.

Using this threshold value, we obtain the $l \times k$ matrix {\bf M}, with matrix elements $M_{cd} = 1$ if country $c$ has a {\textrm RDD} in disease $d$, and zero otherwise. 
A visualization of the matrix {\bf M} that is used to calculate the {\textrm HCI} and the {\textrm DCI} for this dataset is shown in Figure~\ref{fig:bipartite}. 
From this matrix, similar to \citet{hidalgo2009building}, we introduce the {\textrm HCI} as a measure of countries' disease structures.
To obtain the {\textrm HCI}, we first calculate the $l \times l$ square matrix {\bf \~{M}}.
In short, matrix {\bf \~{M}} provides information about links connecting two countries $c$ and $c'$, based on the prevalent cases of diseases in both.
The matrix elements ${\tilde{M}_{cc'}}$ are computed as  
\begin{equation}
 \tilde{M}_{cc'}=\frac{1}{k_{c,0}}\sum_{p}\frac{M_{cd} M_{c'd}}{k_{p,0}},
\end{equation}
where $k_{c,0}=\sum_d M_{cd}$ measures the diversification of country $c$ in terms of its different diseases, and $k_{d,0}=\sum_c M_{cd}$ measures the number of countries with information on prevalent cases of disease $d$.
If {\bf K} is the eigenvector of {\bf \~{M}} associated with the second largest eigenvalue, then according to \citet{hausmann2014atlas}, the {\textrm HCI} is calculated as
\begin{equation}
 {\mathrm{HCI}} = \frac{\mathbf{\textrm K}-\left\langle\mathbf{K}\right\rangle}{{\textrm std}(\mathbf{K})}.
\end{equation}

The {\textrm HCI} reflects the disease-composition of a country's pool of diseases, taking into account the composition of the pools of all other countries. Populations with diseases that have prevalent cases of diseases that occur in many other countries have relatively low health complexity scores, while more health-complex countries have a high prevalence of non-ubiquitous diseases. In other words, a country has a complex disease composition, i.e., it is health-complex, if its diseases have high prevalence in only a few other countries. The {\textrm HCI} is higher for countries with diseases located at the core of the `disease-space' and lower for countries with diseases located at the periphery of the `disease-space'.  
  
\subsection{Disease complexity index}

In a similar manner, but placing the spotlight on diseases rather than countries, we can calculate the \emph{Disease Complexity Index} ({\textrm DCI}). 
In this case, the $k \times k$ matrix {\bf \~{M}} provides information about links connecting two diseases $d$ and $d'$, based on the number of countries in which both diseases have prevalent cases.
Therefore, the matrix elements ${\tilde{M}_{dd'}}$ are computed as  
\begin{equation}
 \tilde{M}_{dd'}=\frac{1}{k_{d,0}}\sum_{c}\frac{M_{cd} M_{cd'}}{k_{c,0}},
\end{equation}
and if {\bf Q} is the eigenvector of {\bf \~{M}} associated with the second largest eigenvalue,
\begin{equation}
 {\mathrm {DCI}} = \frac{\mathbf{\textrm Q}-\left\langle\mathbf{Q}\right\rangle}{{\textrm std}(\mathbf{Q})}.
\end{equation}

\begin{table}[htbp]\caption{List of the five diseases with the highest and lowest {\textrm DCI} values during the period of 1990-2016}
\begin{center}
\begin{adjustbox}{width=\textwidth,totalheight=\textheight,keepaspectratio}
\small
{
\begin{tabular}{cllr}
\toprule
\multicolumn{1}{c}{Code}   &\multicolumn{1}{l}{Disease name}   &\multicolumn{1}{l}{Disease section}   &\multicolumn{1}{c}{DCI}   \\
\midrule
\emph{Highest DCI}\\
459                 &       Malignant skin melanoma &     Neoplasms&      1.210    \\
441           		&     	Colon and rectal cancer &      Neoplasms &      1.170 \\
502             	&      	Peripheral artery disease &      Cardiovascular diseases &      1.169  \\
456          		&     	Pancreatic cancer &     Neoplasms &     1.159   \\
533	       		&       Vascular intestinal disorders &     Neurological disorders &     1.152   \\
                    \\

\emph{Lowest DCI}\\
345             	&   Malaria  &   Neglected tropical diseases and malaria &      -2.045   \\
350                 &   African trypanosomiasis &   Neglected tropical diseases and malaria &      -1.978   \\
370                 &   Maternal obstructed labor and uterine rupture & Maternal and neonatal disorders &      -1.957   \\
358           		&  Yellow fever &  Neglected tropical diseases and malaria & -1.893  \\
340          		&  Tetanus & Other infectious diseases & -1.883  \\
\\

\bottomrule
\end{tabular}
}
\end{adjustbox}
\end{center}
\footnotesize Notes: {\textrm DCI}: Disease Complexity Index; Average values for 1990-2016
\label{tab:0}
\end{table}

As discussed above, the {\textrm HCI} and {\textrm DCI} are computed using in $X_{cd}$ the number of prevalent cases of a disease (according to cause levels 3 or 4) per 100,000 population for 195 countries and for 196 diseases. The time-period covered is from 1990 to 2016. With the age-standardized data (see the discussion in Section \ref{sec: data}), we also calculate the age-standardized health complexity index ({\textrm AHCI}) and the age-standardized disease complexity index ({\textrm ADCI}) by following the same formulas. We use the two indices as alternative measures when checking the robustness of our results. It should be noted here that the computation of the indices is based only on diseases for which a country has a {\textrm RDD} in terms of disease prevalence (the incidence matrix of the bipartite network linking countries to diseases, {\bf M}, reflects whether or not a country has a {\textrm RDD} in a specific disease; see Figure \ref{fig:bipartite}). Table \ref{tab:0} lists the five diseases with the highest and lowest {\textrm DCI} scores averaged over the period of 1990-2016.

\section{The geography of complex diseases}
\label{sec: geography}

Figure \ref{fig:network2} shows the patterns of disease specialization for the world's economies, classified by the World Bank into four income groups - `high', `upper-middle', `lower-middle', and `low'. Diseases in a region where more than half of its countries have a ${\textrm {RDD}} >1$ are shown with black nodes. It seems that high-income countries occupy the core, composed of `non-communicable diseases' such as `pancreatic cancer', `Parkinson disease', `ischemic stroke' and injuries such as `falls', `poisonings` and `other exposure to mechanical forces'. On the other side of the spectrum, low-income countries tend to have a RDD in `communicable, neonatal, maternal and nutritional diseases' that lie in the periphery of the disease space such as `diarrheal diseases', 'encephalitis' and `malaria'. Most of the communicable diseases for which low-income countries have a ${\textrm {RDD}} >1$ also appear in the periphery (for example, `Turner syndrome', `neural tube defects' and `pyoderma'). Examples of injuries for which low-income countries have a RDD include `venomous animal contact' and `sexual violence', which again appear in the periphery of the disease space.    

\begin{figure}[t]
  \centering
 \includegraphics[width = 0.99\textwidth]{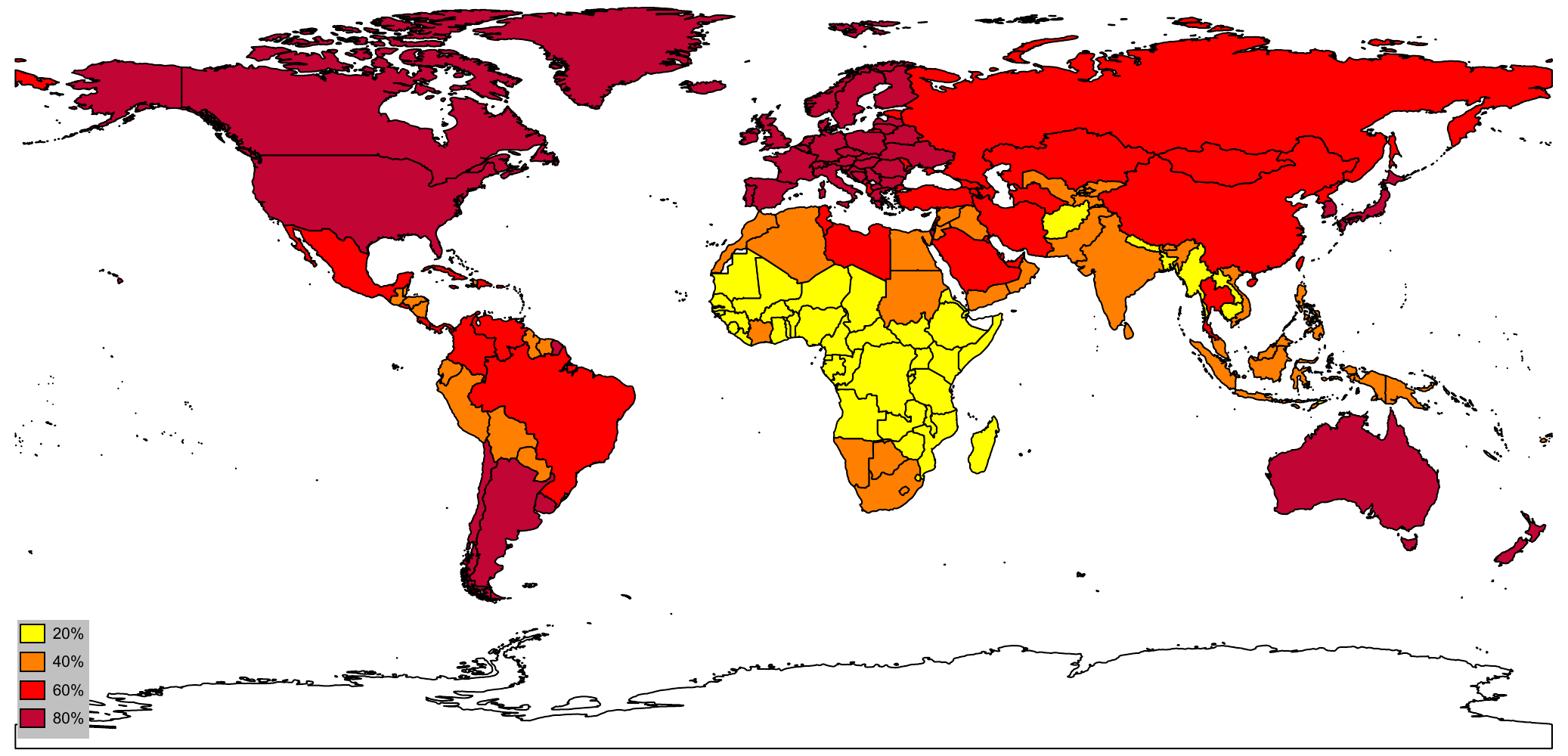}
 \caption{{\bf Health complexity index across the world.} 
\footnotesize Mean values for the period 1990-2016: percentile repartition. Countries depicted in dark red have a mean value above the 80th percentile.}
 \label{fig:map}
\end{figure}

The above descriptive findings are also observable in Figure \ref{fig:map}, where we map the spatial variation in complex diseases. This figure shows the repartition of the {\textrm HCI} across countries when taking average values for the period 1990-2016. We see rather clearly that disease complexity is unevenly distributed in the world and that the most complex countries in terms of diseases seem to be located in Europe, North America, and Australia -- European countries, Australia, the US, and Canada belong to the set of countries with the highest {\textrm HCI} (>80\%). In contrast, most countries in Africa have much lower {\textrm HCIs} on average.

\begin{figure}[t]
\centering
\subfloat[] {\includegraphics[width=15.7cm, height=7.33cm] {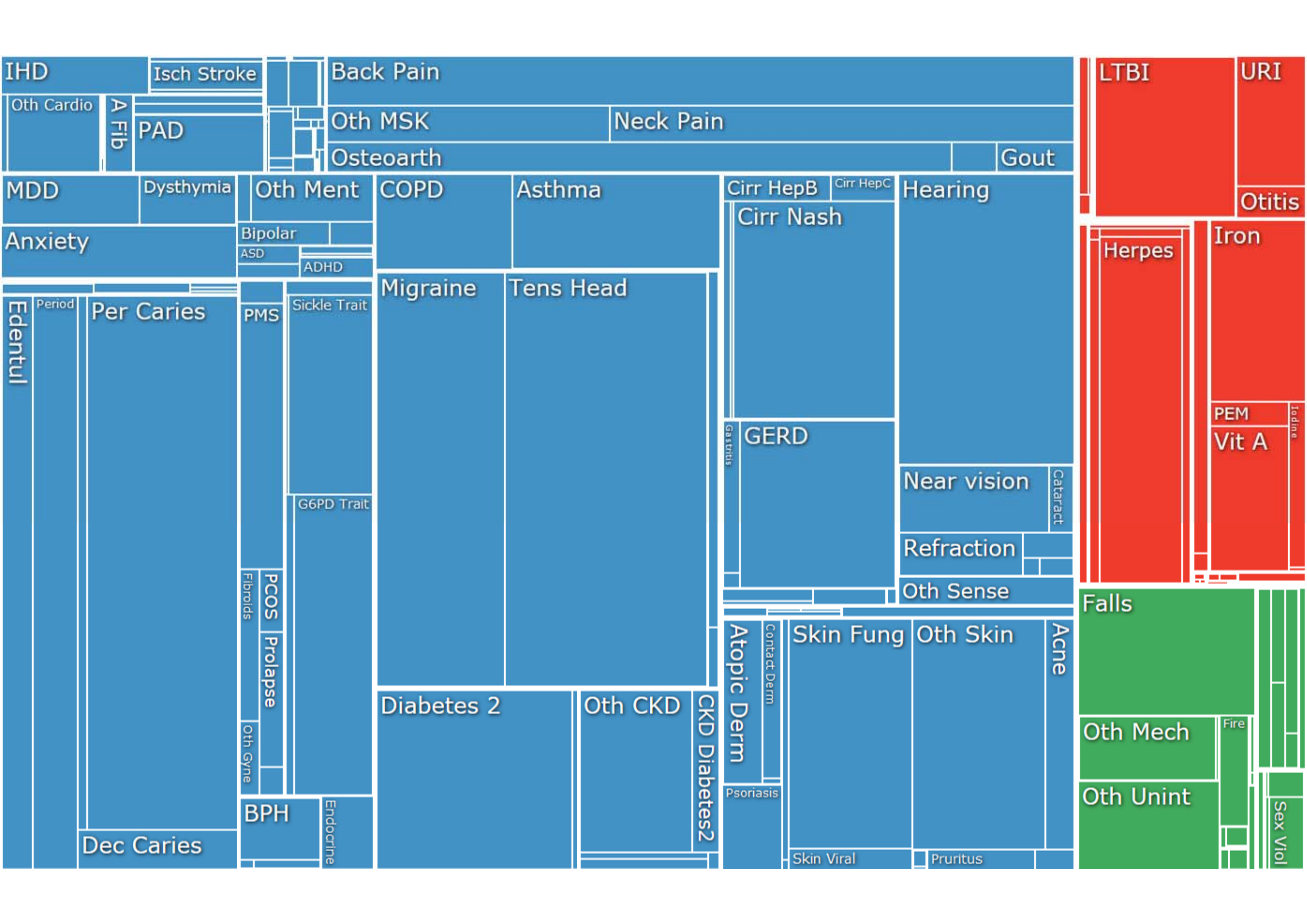}}\\
 \centering
 \subfloat[] {\includegraphics[width=15.7cm, height=7.33cm]{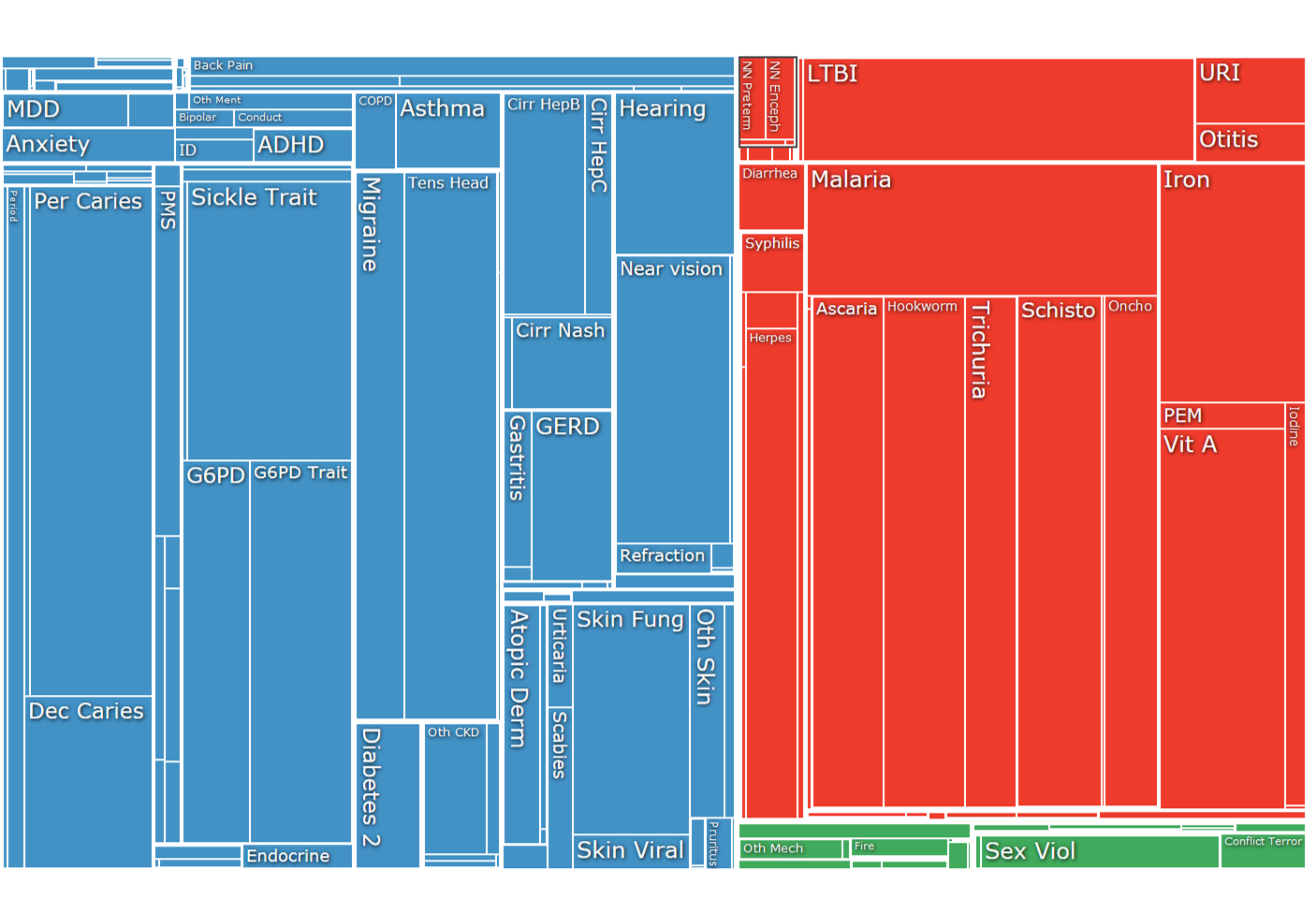}}\\
 \centering
\caption{{\bf The disease maps of two countries.} 
\footnotesize   [a] Portugal  [b] The Democratic Republic of the Congo. Number of prevalent cases per 100,000 population (in 2016) by type of disease: communicable, neonatal, maternal and nutritional diseases (red), non communicable diseases (blue), and injuries (green). Portugal is the country with the highest health complexity index in 2016. The Democratic Republic of the Congo is the least health-complex country in 2016. Source: GBD Compare, IHME Viz Hub, \url{http://www.healthdata.org/results/data-visualizations. }}
\label{fig:portugalcongo}
\end{figure}

Taking a closer look at differences between particular countries, Figure \ref{fig:portugalcongo} displays the disease maps of Portugal and the Democratic Republic of the Congo (DRC), which have the highest and lowest {\textrm HCI} scores (in 2016) respectively. The two countries strongly differ in the composition of their diseases. Portugal has a disease structure in which, out of all prevalent cases of disease, the proportion of non-communicable diseases (97.8\%) is more than double that of communicable diseases (42.4\%). The non-communicable diseases with relatively high proportions of prevalent cases include `tension headache' (37.5\%), `permanent caries' (36.2\%), `migraine' (23.8\%) and `age-related hearing loss' (22.8\%). In the communicable, neonatal, maternal and nutritional diseases category (red), `genital herpes' (13.2\%), `latent TB infection' (10.4\%) and `dietary iron deficiency' (7.9\%) are the diseases with the highest proportion of prevalent cases. Regarding prevalent cases of injuries (green), Portugal's proportion of total cases in the country is 29\%, while the respective rate for the DRC is only 13.6\%.   

The distribution of diseases in the DRC is more uniform: communicable, neonatal, maternal and nutritional diseases comprise 89.5\% of total prevalent cases of diseases, and this figure is comparable to the proportion of non-communicable diseases (93.1\%). For the DRC, `vitamin A deficiency' (23.5\%), 'malaria' (22.9\%) and `schistosomiasis' (21.2\%) are the diseases with the highest proportion of prevalent cases out of all cases of diseases in the country.       

The above `structural' differences are captured by the {\textrm HCI}. Communicable, neonatal, maternal and nutritional diseases are, on average, less complex than non-communicable diseases, because the former lie in the periphery of the disease space, while the latter constitute its core. Hence the DRC receives a lower {\textrm HCI} value compared to Portugal, because in its population, the proportion of communicable, neonatal, maternal and nutritional diseases is higher.

\begin{figure}[t]
  \centering
 \includegraphics[width = 0.65\textwidth]{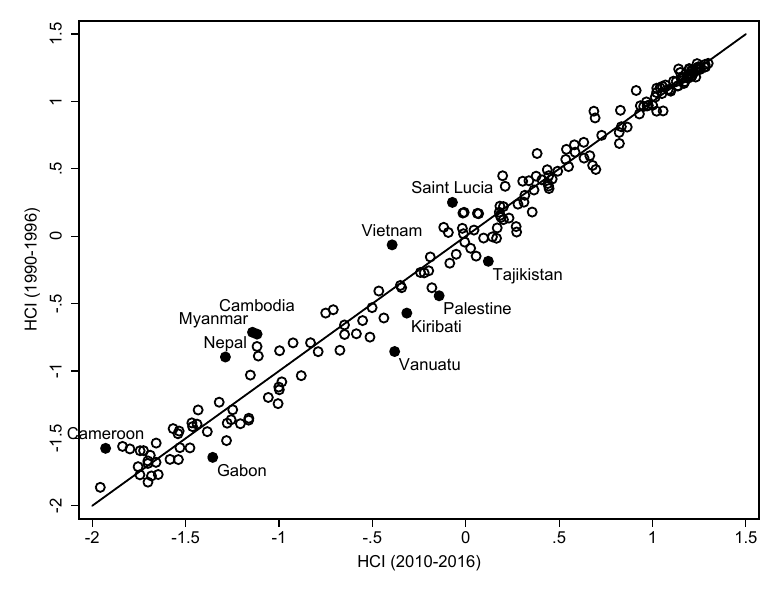}
 \caption{{\bf Evolution of the health complexity index over time. } 
\footnotesize  Changes in the health complexity index from 1990-1996 to 2010-2016. The labelled black dots are the 10 countries with the largest changes over time.}
 \label{fig:graphevol}
\end{figure}

\begin{figure}[t]
  \centering
 \includegraphics[width = 0.99\textwidth]{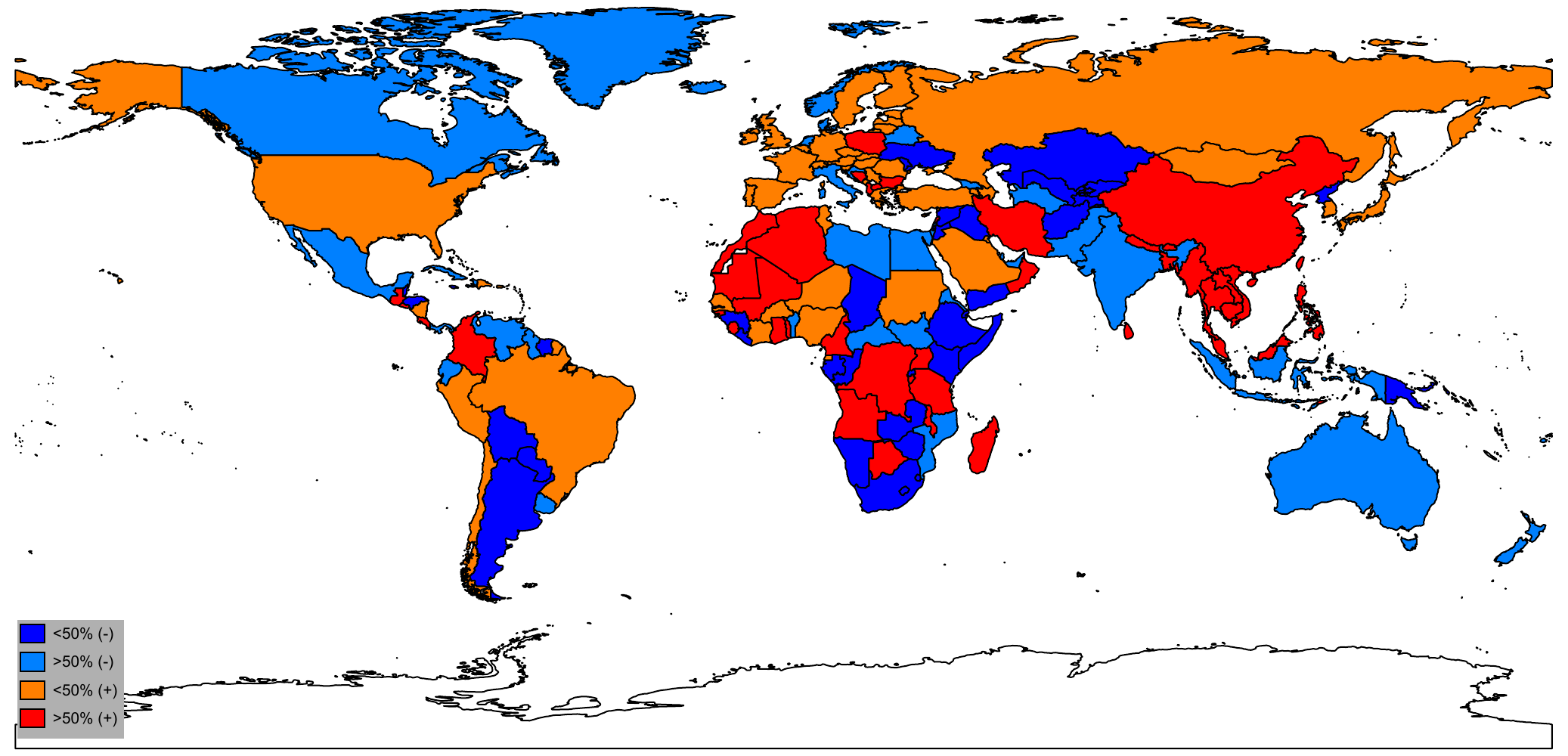}
 \caption{{\bf Changes in the health complexity index across the world. } 
\footnotesize Changes in {\textrm HCI} scores (percentile repartition) from the period 1990-1996 to 2010-2016. Countries depicted in light blue and blue (orange and red) have experienced a decrease (increase) in {\textrm HCI} score below and above the median decrease (increase), respectively. }
 \label{fig:mapevol}
\end{figure}

Regarding the evolution of {\textrm HCI} scores over time, figures \ref{fig:graphevol} and \ref{fig:mapevol} depict how health complexity in our sample of countries has changed from 1990-1996 to 2010-2016. Cambodia, Myanmar, Nepal, Vietnam, Saint Lucia and Cameroon have registered significant increases in the complexity of their diseases. On the other hand, the diseases of countries like Vanuatu, Kiribati, Palestine, Tajikistan and Gabon are now less complex than in the early 1990s. Figure \ref{fig:mapevol} depicts the same information in a world map. Blue and light blue colors depict a decrease in {\textrm HCI} score, while orange and red colors denote countries with an increase in {\textrm HCI} score from 1990-1996 to 2010-2016. From these figures, it can be observed that changes over time are rather small. Hence, it seems that a country's {\textrm HCI} score tends to persist through time, which is to be expected for a metric of prevalent cases of diseases aggregated at the country level. This motivates the inclusion of the lagged value of {\textrm HCI} in the set of explanatory variables when estimating the effect of economic development on health complexity in the next section.

\section{The effect of economic development on health complexity}
\label{sec: econometrics}

We study the effect of economic development on health complexity using data on GDP per capita (from the World Bank's World Development Indicators) and the {\textrm HCI} (see Section \ref{sec: method}).   
Given the availability of controls, the sample covers a minimum of 168 developed and developing countries over the period of 1992-2015. \footnote{Afghanistan,	Albania,	Algeria,	Angola,	Antigua and Barbuda,	Argentina,	Armenia,	Australia,	Austria,	Azerbaijan,	Bahamas,	Bahrain,	Bangladesh,	Barbados,	Belarus,	Belgium,	Belize,	Benin,	Bhutan,	Bolivia,	Bosnia and Herzegovina,	Botswana,	Brazil,	Brunei, Darussalam,	Bulgaria,	Burkina Faso,	Burundi,	Cabo Verde,	Cambodia,	Cameroon,	Canada,	Central African Republic,	Chad,	Chile,	China,	Colombia,	Comoros,	Rep. of the Congo, Costa Rica,	Cote d'Ivoire,	Croatia,	Cyprus,	Czech Republic,	Denmark,	Dominican Republic,	Ecuador,	Arab Rep. of Egypt,	El Salvador,	Equatorial Guinea,	Estonia,	Eswatini,	Ethiopia,	Fiji,	Finland,	France,	Gabon,	Gambia,	Georgia,	Germany,	Ghana,	Greece,	Grenada,	Guatemala,	Guinea,	Guinea-Bissau,	Guyana,	Haiti,	Honduras,	Hungary,	Iceland,	India,	Indonesia,	Islamic Rep. of Iran,	Iraq,	Ireland,	Israel,	Italy,	Jamaica,	Japan,	Jordan,	Kazakhstan,	Kenya,	Kiribati,	Rep. of Korea,	Kuwait,	Kyrgyz Republic,	Lao PDR,	Latvia,	Lebanon,	Lesotho,	Liberia,	Libya,	Lithuania,	Luxembourg,	Macedonia FYR,	Madagascar,	Malawi,	Malaysia,	Maldives,	Malta,	Mauritania,	Mauritius,	Mexico,	Moldova,	Mongolia,	Montenegro,	Morocco,	Mozambique,	Myanmar,	Namibia,	Nepal,	Netherlands,	New Zealand,	Nicaragua,	Niger,	Nigeria,	Norway,	Oman,	Pakistan,	Panama,	Papua New Guinea,	Paraguay,	Peru,	Philippines,	Poland,	Portugal,	Qatar,	Russian Federation,	Rwanda,	Samoa,	Sao Tome and Principe,	Saudi Arabia,	Senegal,	Serbia,	Seychelles,	Sierra Leone,	Singapore,	Slovak Republic,	Slovenia,	South Africa,	Spain,	Sri Lanka,	St. Lucia,	St. Vincent and the Grenadines,	Sudan,	Suriname,	Sweden,	Switzerland,	Tajikistan,	Tanzania,	Thailand,	Togo,	Tonga,	Trinidad and Tobago,	Tunisia,	Turkey,	Uganda,	Ukraine,	United Arab Emirates,	United Kingdom,	United States,	Uruguay,	Vanuatu,	Venezuela RB,	Vietnam,	Rep. of Yemen,	Zambia,	Zimbabwe.}

\subsection{Regression analysis}

Previous research shows that there is a strong positive association between income and indicators of population health such as life expectancy and child mortality. There are various channels through which economic development can stimulate health improvements, for example, via its effect on nutrition (which in turn leads to better resistance to bacterial diseases and faster recovery from illnesses), as well as through greater labor market participation, worker productivity, investment in human capital, investment in public and private health services, savings, fertility, transportation infrastructure, and lifestyle habits. \citep{preston1975changing, smith1999healthy, bloom1998geography, gallup2001economic, who2001, alleyne2002report, bloom2005health, lorentzen2008death, easterlin1999beneficent, hamoudi1999economic}.\footnote{For a review of the empirical evidence see \citep{lange2017effect}.} The term `diseases of affluence' refers to selected diseases and health conditions that are more prevalent in wealthy nations. Examples include mostly non-communicable diseases such as cardiovascular diseases and their nutritional risk factors (overweight and obesity, elevated blood pressure and cholesterol). It has been shown that economic development is a robust predictor of `diseases of affluence' \citep{ezzati2005rethinking, gupta2006rethinking, murray1996global, reddy1998emerging, yusuf2001global}. However, there is also a large and growing literature that investigates the reverse channel, i.e., that better population health leads to economic development \citep{acemoglu2007disease, bleakley2010health}. The argument is that improved health conditions increase population size, which -- in the medium term -- leads to more people entering the labor force, higher capital accumulation, and higher income per capita. 

In order to estimate the effect of economic development on the health complexity of countries we follow a fixed-effects two-stage least squares/instrumental variables (FE 2SLS/IV) strategy, complemented with a difference Generalized Method of Moments (diff-GMM) approach. We regress the baseline specification described by the following equation:

\begin{equation}
 {HCI}_{i,t} = \rho {HCI}_{i,t-1} + \beta_{1} GDPpc_{i,t}+ \beta_{k} controls_{i,t} + \gamma_{i} + \delta_{t} + u_{i,t}.
 \label{eq:econometric}
\end{equation}

Here, the health complexity of country $i$ in period $t$ ($HCI_{i,t}$) depends on the country's level of  economic development in per capita terms (in logs), $GDPpc_{i,t}$. The lagged value of the dependent variable on the right-hand side is included to capture persistence in health complexity. The main variable of interest is $GDPpc$. The parameter $\beta_{1}$ therefore measures the effect of income per capita on health complexity.\footnote{In order to account for possible changes in the relation between economic development and health complexity over the process of economic development, we have experimented with the inclusion of the quadratic specification of GDP per capita in the estimated equation. Our baseline results (which are available upon request) do not confirm a U-shaped relationship between economic development and health complexity.} Additional potential covariates are included in the vector $controls_{i,t}$. The $\gamma_{i}$'s denote a full set of country dummies and the $\delta_{t}$'s denote a full set of time effects that capture common shocks to the health complexity scores of all countries. The error term $u_{i,t}$ captures all other omitted factors, with $E(u_{i,t})=0$ for all $i$ and $t$. To examine the robustness of our results and to generalize our findings, we replicate our analysis for additional/alternative control variables and substitute the {\textrm HCI} with the \textrm{AHCI}, finding qualitatively similar results (see subection \ref{sec:results}).

\begin{table}[htbp]
  \centering
  \caption{Variable definitions, sources and summary statistics}
  \label{tab:summary}
  \begin{adjustbox}{width=\textwidth,totalheight=\textheight,keepaspectratio}%
    \begin{tabular}{p{5.855em}p{22.285em}crr}
    \toprule
    \textbf{Variable} & \textbf{Definition} & \textbf{Source} & \textbf{Mean} & \textbf{Std. Dev.} \\
    \midrule
   HCI   & Health Complexity Index. & Authors' calculations & -0.006 & 1.004 \\\\
   
   HCI$+$  & age-standardized Health Complexity Index. & Authors' calculations  & -0.008 & 1.008 \\\\

   GDPpc & (log) GDP per capita, PPP (constant 2011 international \$) & World Development Indicators & 8.342 & 1.521 \\\\
    
    old & (log) Population aged 65 and above (\% of total) & World Development Indicators & 1.764 & 0.667 \\\\
        
    female & (log) Female population (\% of total)  & World Development Indicators & 3.912 & 0.067 \\\\
        
    urban & (log) Urban population (\% of total)  & World Development Indicators & 3.871 & 0.519 \\\\
    
    agriculture & (log) Agriculture, value added (\% of GDP) & World Development Indicators & 2.054 & 1.228 \\\\
        
    manufacturing & (log) Manufacturing, value added (\% of GDP) & World Development Indicators & 2.407 & 0.591 \\\\
        
    education & (log) Enrollment in secondary education, both sexes (total) & World Development Indicators & 4.229 & 0.587 \\\\
    
    population density & (log) Population density (people per sq. km of land area) & World Development Indicators & 4.123 & 1.398 \\\\
        $CO_2$ & (log)  $CO_2$ emissions (kg per 2010 \$US of GDP)&  World Bank & -0.930 & 0.704 \\\\

    health expenditure & (log)   Total health spending, PPP (thousands of 2017 \$US)  & Global Health Spending 1995-2015, IHME & 15.071 & 2.239 \\\\
    
    economic globalization   & Actual flows (trade, foreign direct investment, stocks, portfolio investment, income payments to foreign nationals), restrictions (hidden import barriers, mean tariff rate, taxes on international trade, capital account restrictions). Higher values reflect greater economic globalization. & KOF Index of Globalization & 54.462 & 16.234 \\\\
    
    political globalization & Embassies in country, membership in international organizations, participation in UN security council missions, international treaties. Higher values reflect greater political globalization. & KOF Index of Globalization & 62.283 & 21.985 \\\\
    
    \bottomrule
    \end{tabular}%
\end{adjustbox}
\end{table}%

\subsubsection{Control Variables}
\label{sec:controls}

We include in the estimated equation a number of control variables that are likely related to health complexity.

The proportion of nations' populations over the age of 65 has been increasing in recent years and will continue to rise in future as a result of longer life expectancy. Population age-structure is a significant determinant of  a nation's health status, due to age-related diseases (i.e., illnesses and conditions that occur more frequently in people as they get older). Examples of age-related diseases include cardiovascular and cerebrovascular diseases, hypertension, cancer, Parkinson's disease, Alzheimer's disease, osteoarthritis and osteoporosis. Demographic factors such as age and sex are considered key covariates in the study of human health and well-being, hence the percentage of $old$ population (aged 65 and above, in logs) and the percentage of $female$ population (in logs) are included in the set of control variables.    

It has previously been shown that sex interacts with social, economic and biological determinants to create different health outcomes for males and females. For example, \citet{vlassoff2007gender} reviews a large number of studies on the interaction between sex and the determinants and consequences of chronic diseases, showing how these interactions result in different approaches to prevention, treatment, and coping with illness.   

In our analysis, we also control for the (log) percentage of urban population, $urban$. According to the World Health Organization (WHO), a  large proportion of non-communicable diseases is linked to risks related to the urban environment, such as physical inactivity and obesity, cardiovascular and pulmonary diseases from transport-generated urban air pollution, ischemic heart disease and cancers from household biomass energy use, asthma from indoor air pollution, and heat-related strokes and illnesses. In addition, communicable diseases such as tuberculosis, dengue fever, and many respiratory and diarrheal diseases result from unhealthy urban environments (e.g. lack of adequate ventilation, unsafe water storage and poor waste management, indoor air pollution, moldy housing interiors, poor sanitation).\footnote{See WHO, Health and sustainable development: About health risks in cities, \url{https://www.who.int/sustainable-development/cities/health-risks/about/en/}}

Industrialization, i.e., the structural transformation from agricultural to industrial production also has a range of significant health implications \citep{mcneill1998plagues, chaudhuri1985trade, steckel2002backbone, szreter2004industrialization, steckel1999industrialization}. We capture these implications in our analysis by including the (log) value added of $agriculture$ (\% of GDP) and the (log) value added of $manufacturing$ (\% of GDP) in the estimated equation.

To check the robustness of our baseline results, we replicate our analysis controlling also for the human capital of the population by utilizing total enrolment in secondary $education$ (in logs). It is well established in the relevant literature that through education, people gain the ability to be effective in their lives. They adopt healthier lifestyles and inspire their offspring to do as well \citep{mirowsky1998education}. Individuals with higher levels of education also tend to have better socioeconomic resources for a healthy lifestyle and a higher probability of living and working in healthy environments \citep{adler2002socioeconomic, braveman2010socioeconomic}. In addition, educated individuals tend to have lower exposure to chronic stress \citep{pampel2010socioeconomic}. Low educational attainment, on the other hand, is associated with a shorter life expectancy, poor self-reported health, and a high prevalence of infectious and chronic non-infectious diseases \citep{feldman1989national, guralnik1989predictors, liu1982relationship, woodward1992social}.\footnote{\citet{grossman2006education} and \citet{ross1995links} review the relationships between education and a wide variety of health measures.} Furthermore, we re-estimate the baseline model by substituting $urban$ with $population$ $density$ (people per square km of land, in logs).

The variable $CO_2$ (log of $CO_2$ emissions in kg per 2010 \$US of GDP) captures the effect of air pollution on health, which has been the subject of numerous studies in recent years (for an extensive review, see \citep{brunekreef2002air}).      

Finally, $health$ $expenditure$ (log of total health spending in thousands of purchasing power parity (PPP)-adjusted 2017 \$US) is also included in the set of explanatory variables controlling for the association between healthcare spending and health outcomes \citep{werblow2007population, barlow1999determinants, cremieux1999health, wolfe1986health, mackenbach1991health, seshamani2004ageing, zweifel1999ageing, nixon2006relationship} 

Data definitions, sources and summary statistics for the variables included in the analysis are given in Table \ref{tab:summary}.

\subsubsection{Instrumental variables}
\label{sec:instruments}

We estimate equation (\ref{eq:econometric}) using different econometric methods. First, we use fixed-effects OLS. However, fixed effects estimators do not necessarily identify the effect of economic development on health complexity. The estimation of causal effects requires exogenous sources of variation. While we do not have an ideal source of exogenous variation recognized by previous studies, there are two promising potential instruments of {\textrm economic development} that we adopt in our fixed-effects 2SLS/IV and diff-GMM analyses. 

First, we use the KOF Swiss Economic Institute's $economic$ $globalization$ index, characterized as the flows of goods, capital, and services, as well as information and perceptions that accompany market exchanges \citep{dreher2006does}. Higher values reflect greater economic globalization.

The second instrument considered is the KOF Swiss Economic Institute's $political$ $globalization$ index, characterized by the number of embassies in a country, its membership in international organizations, its participation in UN security council missions and international treaties. Higher values reflect greater political globalization.

There is extensive research documenting the positive relationship between globalization and economic development and growth \cite{dreher2006does, gozgor2017causal, duttagupta2018globalization, gurgul2014globalization, chang2010globalization}. While we do not have a precise theory to support the prediction, it is expected that changes in the $economic$ $globalization$ and $political$ $globalization$ indices have no direct effect on a country's disease structure and impact health complexity only indirectly, through the channel of economic development. This point is also verified in our dataset, as we find no correlation between {\textrm HCI} scores and these two variables (the results are available upon request).

\begin{table}[htbp]
  \centering
  \caption{The effect of economic development on health complexity: Fixed-effects OLS}
  \label{tab:t2}
   \begin{adjustbox}{width=\textwidth,totalheight=\textheight,keepaspectratio}%
\begin{tabular}{p{12.20em}*{6}{c}}
\toprule
                    &\multicolumn{1}{c}{(1)}   &\multicolumn{1}{c}{(2)}   &\multicolumn{1}{c}{(3)}   &\multicolumn{1}{c}{(4)}   &\multicolumn{1}{c}{(5)} &\multicolumn{1}{c}{(6)} \\
\midrule
    $HCI_{t-1}$ & 0.938*** & 0.938*** & 0.938*** & 0.933*** & 0.929*** & 0.929*** \\
          & (0.009) & (0.009) & (0.010) & (0.010) & (0.010) & (0.010) \\
    $GDPpc$ & 0.013*** & 0.013*** & 0.013*** & 0.012*** & 0.009** & 0.011*** \\
          & (0.003) & (0.003) & (0.003) & (0.003) & (0.003) & (0.004) \\
    $old$ &       & -0.001 & -0.001 & 0.002 & 0.000 & 0.002 \\
          &       & (0.006) & (0.006) & (0.006) & (0.007) & (0.007) \\
    $female$ &       &       & -0.014 & -0.013 & 0.002 & -0.006 \\
          &       &       & (0.024) & (0.024) & (0.023) & (0.024) \\
    $urban$ &       &       &       & 0.020** & 0.025*** & 0.021** \\
          &       &       &       & (0.008) & (0.008) & (0.009) \\
    $agriculture$ &       &       &       &       & -0.006*** & -0.006*** \\
          &       &       &       &       & (0.002) & (0.002) \\
    $manufacturing$ &       &       &       &       &       & 0.001 \\
          &       &       &       &       &       & (0.002) \\
    Observations & 4,750 & 4,593 & 4,593 & 4,593 & 4,377 & 4,181 \\
    Countries & 190   & 183   & 183   & 183   & 180   & 177 \\
    R-squared & 0.873 & 0.874 & 0.874 & 0.875 & 0.872 & 0.872 \\
\bottomrule
\multicolumn{7}{c}{%
\begin{minipage}{17.9cm}%
Note: Dependent variable: Health Complexity Index ({\textrm HCI}). Main independent variable: GDP per capita in logs ($GDPpc$). Time fixed-effects are included in all regressions. Robust standard errors are in parentheses. * p<0.10, ** p<0.05, *** p<0.01 
\end{minipage}}
\end{tabular}
\end{adjustbox}
\end{table}

\subsection{Regression results}
\label{sec:results}

In this section, we discuss the results of estimating equation (\ref{eq:econometric}) with different econometric techniques. Table \ref{tab:t2} reports the results of fixed-effects ordinary least squares (FE-OLS) with time dummies, adding an additional variable from the set of controls in each step (column). In all specifications, economic development has a positive relationship with health complexity, and the control variables enter with the expected sign. The $agriculture$ coefficient is negative, and countries with a higher proportion of $urban$ population exhibit greater health complexity. 

The results depicted in Table \ref{tab:t2} could only be interpreted as correlations. For a robust analysis accounting for the potential endogeneity problem in the relationship under consideration (discussed above), we use fixed-effects 2SLS/IV estimation techniques complemented with diff-GMM estimations \`a la Arellano-Bond \citep{arellano1991some}. Table \ref{tab:t3} presents our baseline results. 

In columns (1)-(8), we estimate equation~(\ref{eq:econometric}) with FE 2SLS/IV regressions. We use time dummies and robust standard errors (in parentheses). In all cases, $GDPpc$ is a positive and statistically significant predictor of health complexity. In fact, we find that an increase in $GDP$ $per$ $capita$ of 10\% is associated with an improvement of about 0.003 in the {\textrm HCI} (standard deviation: 1.004). This positive impact of economic development on health complexity is robust to the inclusion of control measures discussed in subsection~\ref{sec:controls}. The statistically significant $urban$ coefficient implies that a higher proportion of urban population is associated with more complex diseases. 

In the fixed effects 2SLS/IV estimations we report: (a) the $F$-$test$ for the joint significance of the instruments in the first stage: the rule of thumb is to exceed 10, hence the test implies weak significance \citep{16284}; (b) the Durbin-Wu-Hausmann ($DWH$) test for the endogeneity of regressors: the null hypothesis that the IV regression is not required is rejected; (c) the Cragg-Donald F-statistic ($Weak$-$id$), testing the relevance of the instruments in the first-stage regression: no evidence of a low correlation between instruments and the endogenous regressor is found after controlling for the exogenous regressors; (d) the  Kleibergen-Paap Wald test ($LM$-$weakid$) of weak identification: the null hypothesis that the model is weakly identified is rejected; (e) the p-value for Hansen's test of overidentification: the acceptance of the null indicates that the overidentifying restrictions cannot be rejected. 

In column (9) of Table \ref{tab:t3} we report the diff-GMM estimations including year fixed effects and robust standard errors. The results verify the previous findings both qualitatively and quantitatively, i.e., the estimated coefficient of $GDPpc$ implies an improvement of 0.003 in the {\textrm HCI} with a $GDP$ $per$ $capita$ increase of 10\%. Among the control variables, only the $urban$ variable has a statistically significant and positive sign. The values reported for AR(1) and AR(2) are the p-values for first- and second-order autocorrelated disturbances. As expected, there is high first-order autocorrelation and no evidence for significant second-order autocorrelation. Hence, our test statistics hint at a proper specification.

\newpage
\begin{landscape}
{
\begin{table}[htbp]
  \centering
\caption{The effect of economic development on health complexity: baseline results}
\label{tab:t3}
\begin{adjustbox}{max width=1.22\textwidth}%
\begin{tabular}{l*{9}{c}}
\toprule
                    &(1)&(2)&(3)&(4)&(5)&(6)&(7)&(8)&(9)\\
                    & FE 2SLS/IV & FE 2SLS/IV & FE 2SLS/IV & FE 2SLS/IV & FE 2SLS/IV & FE 2SLS/IV & FE 2SLS/IV & FE 2SLS/IV & diff-GMM\\
\midrule
$HCI_{t-1}$   & 0.926*** & 0.927*** & 0.927*** & 0.923*** & 0.923*** & 0.922*** & 0.922*** & 0.921*** & 0.814*** \\
          & (0.010) & (0.010) & (0.010) & (0.010) & (0.010) & (0.010) & (0.011) & (0.010) & (0.061) \\
$GDPpc$ & 0.031*** & 0.031*** & 0.031*** & 0.031*** & 0.026** & 0.029*** & 0.030* & 0.027** & 0.032** \\
          & (0.011) & (0.011) & (0.011) & (0.011) & (0.011) & (0.011) & (0.017) & (0.013) & (0.013) \\
$old$   &       & -0.003 & -0.002 & -0.001 & -0.003 & -0.001 & -0.002 & -0.001 & -0.005 \\
          &       & (0.005) & (0.005) & (0.005) & (0.006) & (0.006) & (0.006) & (0.006) & (0.017) \\
$female$ &       &       & -0.015 & -0.015 & -0.019 & -0.029 & -0.031 & -0.026 & 0.029 \\
          &       &       & (0.023) & (0.023) & (0.027) & (0.028) & (0.033) & (0.029) & (0.044) \\
$urban$ &       &       &       & 0.013 & 0.020*** & 0.017** & 0.017** & 0.017** & 0.031* \\
          &       &       &       & (0.008) & (0.008) & (0.008) & (0.009) & (0.008) & (0.018) \\
$agriculture$ &       &       &       &       & -0.003 & -0.003 & -0.002 & -0.003 & -0.002 \\
          &       &       &       &       & (0.003) & (0.003) & (0.004) & (0.003) & (0.008) \\
$manufacturing$ &       &       &       &       &       & 0.002 & 0.002 & 0.002 & -0.003 \\
          &       &       &       &       &       & (0.002) & (0.002) & (0.002) & (0.012) \\
\midrule
Fist-stage results &       &       &       &       &       & & & &  \\
$economic$ $globalization$ & 0.003*** & 0.003*** & 0.003*** & 0.003*** & 0.004*** & 0.003*** & 0.005*** &  \\
          & (0.001) & (0.001) & (0.001) & (0.001) & (0.000) & (0.000) & (0.000) &  \\
$political$ $globalization$ & 0.004*** & 0.004*** & 0.004*** & 0.004*** & 0.004*** & 0.005*** &       & 0.006*** \\
          & (0.001) & (0.001) & (0.001) & (0.001) & (0.000) & (0.000) &       & (0.000) \\
	\midrule
    Observations & 4,168 & 4,143 & 4,143 & 4,143 & 3,976 & 3,826 & 3,826 & 3,926 & 3,652 \\
    Countries & 171   & 170   & 170   & 170   & 170   & 168   & 168   & 173   & 168 \\
    F-test & 37.24 & 37.05 & 36.31 & 39.93 & 56.26 & 61.41 & 79.18 & 76.56 &  \\
    DWH-test & 2.978 & 2.911 & 2.785 & 3.073 & 2.500 & 2.646 & 1.304 & 1.523 &  \\
    Weak-id & 59.73 & 54.25 & 54.88 & 56.82 & 97.43 & 107.0 & 104.3 & 152.6 &  \\
    LM-weakid & 71.12 & 70.95 & 69.68 & 75.01 & 109.7 & 116.3 & 72.90 & 73.98 &  \\
    Hansen (p-value) & 0.590 & 0.699 & 0.676 & 0.897 & 1.000 & 0.904 &       &       &  \\
    AR(1) &       &       &       &       &       &       &       &       & 0.000 \\
    AR(2) &       &       &       &       &       &       &       &       & 0.475 \\
\bottomrule
\multicolumn{10}{l}{%
\begin{minipage}{26.1cm}%
\footnotesize Note: Dependent variable: Health Complexity Index ({\textrm HCI}). Main independent variable: GDP per capita in logs ($GDPpc$). Columns (1)-(8): Fixed-effects 2SLS/IV; {\textrm HCI} is instrumented: To save space, we only include the first-stage estimated coefficients of the instruments. The results for the rest of the variables are available upon request. Column (9): One-step diff-GMM. All regressions include time dummies. Robust standard errors are in parentheses. F-test gives the F-statistic for the joint significance of the instruments in the first stage. DWH-test is the Durbin-Wu-Hausman test of endogeneity of the regressors. Weak-id gives the Cragg-Donald F-statistic for weak identification. LM-weakid gives the Kleibergen-Paap Wald test of weak identification. Hansen (p-value) gives the p-value of the Hansen test of overidentification. AR(1) and AR(2) are the p-values for first- and second-order autocorrelated disturbances.  * p<0.10, ** p<0.05, *** p<0.01
\end{minipage}}
\end{tabular}
\end{adjustbox}
\end{table}
}
\end{landscape}

\begin{table}[htbp]
  \centering
  \caption{The effect of economic development on health complexity: \textrm{AHCI}}
  \label{tab:t4}
   \begin{adjustbox}{width=\textwidth,totalheight=\textheight,keepaspectratio}%
\begin{tabular}{p{17.20em}*{4}{c}}
\toprule
                    &\multicolumn{1}{c}{(1)}   &\multicolumn{1}{c}{(2)}   &\multicolumn{1}{c}{(3)}   &\multicolumn{1}{c}{(4)}    \\
                    & FE 2SLS/IV & FE 2SLS/IV & FE 2SLS/IV & diff-GMM \\
\midrule
    $AHCI_{t-1}$ & 0.923*** & 0.925*** & 0.919*** & 0.828*** \\
          & (0.011) & (0.012) & (0.011) & (0.114) \\
    $GDPpc$ & 0.037*** & 0.033** & 0.041*** & 0.024* \\
          & (0.011) & (0.015) & (0.014) & (0.013) \\
    $female$ & -0.028 & -0.021 & -0.037 & 0.030 \\
          & (0.027) & (0.034) & (0.031) & (0.040) \\
    $urban$ & 0.009 & 0.010 & 0.007 & 0.011 \\
          & (0.007) & (0.008) & (0.008) & (0.023) \\
    $agriculture$ & 0.002 & 0.001 & 0.003 & 0.008 \\
          & (0.003) & (0.004) & (0.004) & (0.010) \\
    $manufacturing$ & 0.002 & 0.002 & 0.003 & -0.013 \\
          & (0.002) & (0.002) & (0.002) & (0.011) \\
\midrule
    Fist-stage results &       &       &       &       \\
    $economic$ $globalization$ & 0.004*** & 0.005*** & &  \\
          & (0.000) & (0.000) &  &   \\
    $political$ $globalization$ & 0.004*** &  & 0.005*** &  \\
          & (0.000) &  & (0.000) &  \\
	\midrule
    Observations & 3,826 & 3,826 & 3926 & 3,652 \\
    Countries & 168   & 168   & 173   & 168 \\
    F-test & 67.39 & 101.5 & 71.22 &  \\
    DWH-test & 6.055 & 2.200 & 4.548 &  \\
    Weak-id & 116.2 & 136.4 & 138.9 &  \\
    LM-weakid & 125.8 & 92.43 & 67.20 &  \\
    Hansen (p-value) & 0.681 &       &       &  \\
    AR(1) &       &       &       & 0.000 \\
    AR(2) &       &       &       & 0.158 \\
\bottomrule
\multicolumn{5}{c}{%
\begin{minipage}{16.8cm}%
Note: Dependent variable: Age-standardized Health Complexity Index (\textrm{AHCI}). Main independent variable: GDP per capita in logs ($GDPpc$). Columns (1)-(3): Fixed-effects 2SLS/IV; $AHCI$ is instrumented: To save space, we only include the first-stage estimated coefficients of the instruments. The results for the rest of the variables are available upon request. Column (4): One-step diff-GMM. All regressions include time dummies. Robust standard errors are in parentheses. F-test gives the F-statistic for the joint significance of the instruments in the first stage. DWH-test is the Durbin-Wu-Hausman test of endogeneity of the regressors. Weak-id gives the Cragg-Donald F-statistic for weak identification. LM-weakid gives the Kleibergen-Paap Wald test of weak identification. Hansen (p-value) gives the p-value of the Hansen test of overidentification. AR(1) and AR(2) are the p-values for first- and second-order autocorrelated disturbances.  * p<0.10, ** p<0.05, *** p<0.01
\end{minipage}}
\end{tabular}
\end{adjustbox}
\end{table}

\newpage
{
\begin{table}[htbp]
  \centering
\caption{The effect of economic development on health complexity: robustness checks}
\label{tab:t5}
\begin{adjustbox}{max width=1.22\textwidth}%
\begin{tabular}{l*{5}{c}}
\toprule
                    &(1)&(2)&(3)&(4)&(5)\\
                    & diff-GMM & diff-GMM & diff-GMM & diff-GMM & diff-GMM \\
\midrule
$HCI_{t-1}$   & 0.752*** & 0.871*** & 0.812*** & 0.920*** & 0.666*** \\
          & (0.059) & (0.046) & (0.063) & (0.040) & (0.062) \\
$GDPpc$ & 0.031* & 0.031*** & 0.024* & 0.042*** & 0.038* \\
          & (0.016) & (0.011) & (0.013) & (0.016) & (0.020) \\
$old$  & -0.022 & -0.010 & 0.004 & -0.002 & -0.005 \\
          & (0.024) & (0.017) & (0.018) & (0.016) & (0.029) \\
$female$ & 0.053 & 0.018 & 0.053 & -0.068 & -0.170 \\
          & (0.050) & (0.051) & (0.045) & (0.047) & (0.124) \\
$urban$ & 0.055 &       & 0.013 & 0.022        &  \\
          & (0.035) &       & (0.017) & (0.023)        &  \\
$agriculture$ & 0.007 & -0.001 & 0.003 & 0.001 & 0.015 \\
          & (0.009) & (0.006) & (0.009) & (0.008) & (0.011) \\
$manufacturing$ & 0.008 & -0.003 & -0.015 & 0.014 & 0.020 \\
          & (0.012) & (0.011) & (0.010) & (0.011) & (0.014) \\
$education$ & -0.003 &       &             &        & 0.026 \\
          & (0.011) &       &            &      & (0.017) \\
$population$ $density$ &       & 0.000 &              &        & -0.073 \\
          &       & (0.016) &             &       & (0.049) \\
$CO_2$   &       &       & 0.013*        &        & -0.003 \\
          &       &       & (0.007)        &       & (0.015) \\
$health$ $expenditure$ &       &       &             & -0.006        & -0.012 \\
          &       &       &              & (0.012)        & (0.018) \\
	\midrule
    Observations & 2,347 & 3,642 & 3,470 & 3,158 & 1,936 \\
    Countries & 161   & 168   & 168   & 168   & 158 \\
    AR(1) & 0.000     & 0.000 & 0.000   & 0.000 & 0.000 \\
    AR(2) & 0.395 & 0.505 & 0.621 & 0.525 & 0.274 \\
\bottomrule
\multicolumn{6}{l}{%
\begin{minipage}{14.3cm}%
\footnotesize Note: Dependent variable: Health Complexity Index ({\textrm HCI}). Main independent variable: GDP per capita in logs ($GDPpc$). Regression analysis: One-step diff-GMM. All regressions include time dummies. Robust standard errors are in parentheses. AR(1) and AR(2) are the p-values for first- and second-order autocorrelated disturbances.  * p<0.10, ** p<0.05, *** p<0.01
\end{minipage}}
\end{tabular}
\end{adjustbox}
\end{table}
}

In Tables \ref{tab:t4} and \ref{tab:t5}, we investigate the robustness of our baseline findings. First, we substitute the {\textrm HCI} with the age-standardized health complexity index (\textrm {AHCI}) maintaining the same set of controls (and time dummies) as in the baseline specification. Second, we investigate whether the positive impact of economic development on health complexity persists under additional and/or alternative control measures (including time dummies). In all cases, the baseline results remain qualitatively intact. In particular, the coefficient of $GDPpc$ is positive and statistically significant in the instrumented regressions (see Table \ref{tab:t4}; to save space, we only include the first-stage estimated coefficients of the instruments -- the results for the rest of the variables are available upon request).

Table \ref{tab:t5} starts from the baseline specification with the full set of controls [column (9) in Table \ref{tab:t3}] and introduces additional variables or alternative measures for some of the previous controls. Specifically, in column (1), we add $education$ (enrolment in secondary education in logs). In column (2), we substitute the $urban$ population variable with $population$ $density$ (people per sq. km of land in logs). In columns (3) and (4), we employ (log) $CO_2$ emissions ($CO_2$ emissions, kg per 2010 \$US of GDP) and (log) $health$ $expenditure$ (total health spending, thousands of 2017 PPP adjusted \$US), respectively. Finally, in column (5), we consider all of the above variables together. Adding these controls in our estimations leaves the findings qualitatively and quantitatively intact. 

The above analysis suggests that economically developed countries tend to exhibit more complex disease structures. Furthermore, exploiting the temporal variation in the data, the fixed-effects 2SLS/IV analysis and the difference GMM estimators reveal a positive, statistically significant, and robust impact of economic development on health complexity. 

\section{Economic development and disease complexity}
\label{sec: economic}

The economic complexity methodology provides a useful toolbox that allows us to compute indices that quantify the complexity of both countries and diseases. 
For example, using the same methodology that computes the {\textrm HCI}, but placing the spotlight on diseases rather than countries, we calculate the {\textrm DCI} (see Section \ref{sec: method}). 
This index quantifies the complexity of countries' diseases according to their prevalent cases worldwide.
Using the economic complexity methodology, \citet{hartmann2017linking} recently introduced a measure that associates products with income inequality and showed how the development of new products is associated with changes in income inequality. 
Here, to decompose economic development at the disease level, we introduce a measure that links a disease to the average income per capita of the countries in which the disease has prevalent cases i.e., an estimate of the expected income per capita related to different diseases.  
In this way, we illustrate how disease complexity is being affected by the level of economic development and quantify the relationship between countries' income per capita and the complexity of their diseases.

Following the methodology in \citet{hartmann2017linking}, we define the \emph{Disease-Income Complexity Index} ({\textrm DICI}), and decompose the relationship between the {\textrm DCI} and the {\textrm DICI} for the prevalent cases of diseases in our sample of countries.\footnote{We also computed the {\textrm ADICI} and investigated its relationship with the {\textrm ADCI}, finding similar results.}

\begin{figure}[t]
  \centering
 \includegraphics[width = 0.85\textwidth]{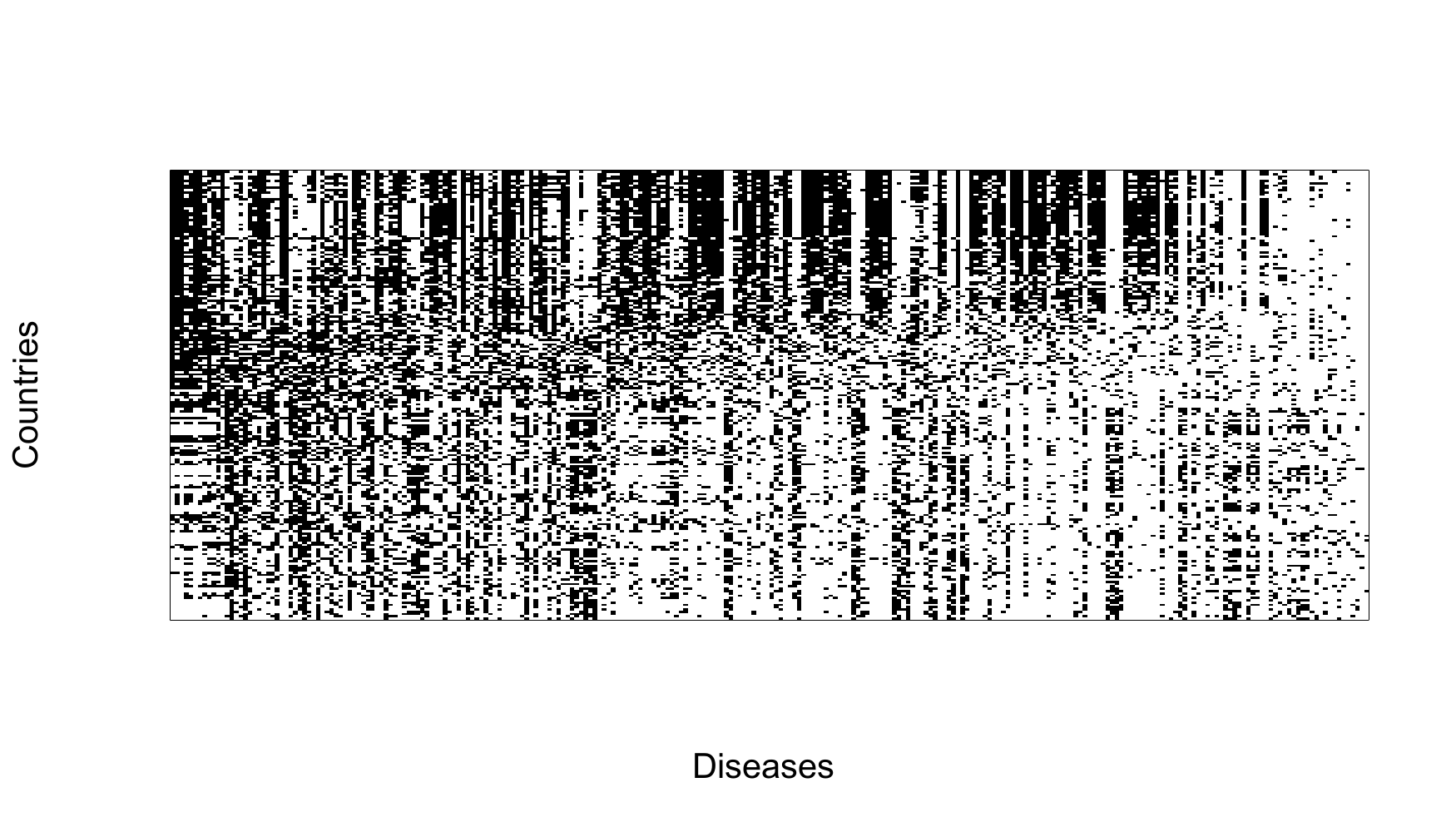}
 \caption{{\bf Matrix representation of the links between countries and diseases.} 
\footnotesize A visualization of this matrix for the year 2016, where 
 a dark point indicates that country $c$ has a {\textrm RDD} in a given disease $d$. 
 The matrix is sorted using the NODF algorithm~\citep{almeida2008consistent}, which highlights the existence of countries that are very well diversified and countries that have prevalent cases only for a small set of diseases.}
 \label{fig:bipartite}
\end{figure}

\subsection{Disease-income complexity index}

Assuming that we have information for $l$ countries and $k$ diseases, we can fill the $(l \times k)$ matrix {\bf M} so that its matrix element $M_{cd} = 1$ if country $c$ has a {\textrm RDD} in disease $d$, and zero otherwise (see Section \ref{sec: method}).
Our dataset contains information for 195 developed and developing countries and for 196 diseases from 1990 to 2016.
A visualization of the matrix {\bf M} that is used to calculate the {\textrm HCI} and the {\textrm DCI} for this dataset is shown in Figure~\ref{fig:bipartite}. 

Every disease $d$ can have prevalent cases in a country $c$. For every disease $d$, we can calculate the fraction $s_{cd}$:
\begin{equation}
 s_{cd}=\frac{X_{cd}}{\sum_{d'}X_{cd'}},
\end{equation}
where $X_{cd}$ is the number of prevalent cases per 100,000 population for disease $d$ in country $c$, while $\sum_{d'}X_{cd'}$ is the number of prevalent cases of all diseases in country $c$.
If $GDP_c$ is the (log) $GDP$ $per$ $capita$ of country $c$, we can calculate the ${\textrm {DICI}}_d$ for every disease $d$ as:
\begin{equation}
 {\textrm {DICI}}_d = \frac{1}{N_d}\sum_c M_{cd}s_{cd}GDP_c,
\end{equation}
where $N_d=\sum_c M_{cd}s_{cd}$ is a normalization factor.

\begin{figure}[t]
 \centering
 \includegraphics[width = 0.49\textwidth]{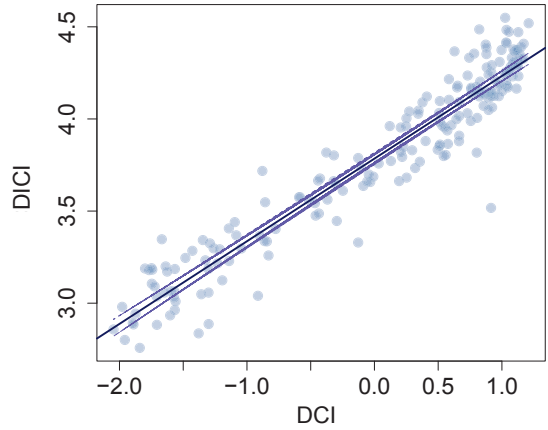}
 \caption{{\bf DICI against DCI.} \footnotesize The solid line represents the fit of a linear model and the dashed line a 95\% prediction interval based on the fitted linear model.}
 \label{fig:cor}
\end{figure}

\begin{table}[htbp]\caption{List of the five diseases with the highest and lowest {\textrm DICI} values during the period of 1990-2016}
\begin{center}
\begin{adjustbox}{width=1.00\textwidth}
\small
{
\begin{tabular}{cllrr}
\toprule
\multicolumn{1}{c}{code}   &\multicolumn{1}{l}{disease name}   &\multicolumn{1}{l}{disease section}   &\multicolumn{1}{c}{DICI}   &\multicolumn{1}{c}{DCI}  \\
\midrule
\emph{Highest DICI}\\
554                 &       Motor neuron disease &     Neurological disorders&      4.55  & 1.025  \\
                   
459           		&     	Malignant skin melanoma &      Neoplasms &      4.52& 1.210 \\
                   
573             	&      	Anorexia nervosa &      Mental disorders &      4.49& 0.818  \\
    
483          		&     	Mesothelioma &     Neoplasms &     4.49 & 1.041  \\
                   
485	       		    &       Non-Hodgkin's lymphoma &     Neoplasms &     4.48& 1.057   \\
                    \\
\emph{Lowest DICI}\\

404           		&   Acute hepatitis E &  Other infectious diseases & 2.89 & -1.707 \\

345             	&   Malaria &   Neglected tropical diseases and malaria &      2.86 & -2.045  \\

353                 &   Cystic echinococcosis &   Neglected tropical diseases and malaria &      2.84 & -1.383  \\

370                 &   Maternal obstructed labor and uterine rupture & Maternal and neonatal disorders &      2.80  & -1.957 \\

359           		&  Rabies &  Neglected tropical diseases and malaria & 2.76 & -1.840 \\

                    \\                    
         
\bottomrule
\end{tabular}
}
\end{adjustbox}
\end{center}
\footnotesize Notes: {\textrm DICI}: Disease-Income Complexity Index; {\textrm DCI}: Disease Complexity Index; Average values for 1990-2016
\label{tab:6}
\end{table}

The {\textrm DICI} is defined at the disease level as the average level of (log) $GDP$ $per$ $capita$ of the countries that have a {\textrm RDD} in disease $d$, weighted by the disease's importance in each country's pool of diseases. Utilizing the (log) PPP $GDP$ $per$ $capita$ (constant 2011 international \$) from the World Bank's World Development Indicators for the countries in our sample, we calculate the above index for every year in the period of 1990-2016. 

Table \ref{tab:6} lists the five diseases with the highest and lowest average {\textrm DICI} values during the period of 1990-2016. It is evident that higher economic development is associated with more complex diseases such as motor neuron disease and malignant skin melanoma. At the other end of the spectrum, less complex diseases such as acute hepatitis E and malaria are associated with low levels of income per capita.  

\subsection{Linking disease complexity and economic development}

In this subsection, we test the existence of a bivariate relationship between the {\textrm DCI} and the {\textrm DICI}. 
Thus, we calculate Pearson's correlation coefficient for {\textrm DICI} against {\textrm DCI}. 
If such an association exists, it should allow us to derive expectations about whether disease complexity can be associated with economic development and verify, with disease-level data, the statistically significant and positive relationship between health complexity and economic development that we found above (Section \ref{sec:results}).
The correlation coefficient for the relationship between the average values of the {\textrm DICI} and the {\textrm DCI} for the period of 1990-2016 is $\rho=0.96\pm{0.01}$ with a p-value $< 2.2\times10^{-16}$.
In Figure~\ref{fig:cor}, we present the scatter-plot of the relationship between the {\textrm DICI} and the {\textrm DCI} for the 196 diseases in our dataset (average values for 1990-2016), together with the fitted linear model.
The slope of the linear fit is the corresponding correlation coefficient.

The statistically significant positive correlation between the {\textrm DICI} and the {\textrm DCI} indicates that more complex diseases are associated with more developed countries, as measured by the (log) $GDP$ $per$ $capita$. This allows us to understand which sets of diseases are linked to better overall economic performance, based on their complexity.

In Table \ref{tab:t7}, we run panel regressions between the {\textrm DCI} and {\textrm DICI}. The results show that the relationship between the {\textrm DCI} and the {\textrm DICI} is the outcome of the correlations both \textit{between} diseases (regression on group means) and \textit{within} diseases (fixed-effects regression with time dummies and standard errors adjusted for disease clusters). 
This suggests that the positive effect of economic development on the complexity of diseases is due to both changes in the structure of the disease space towards more complex diseases and increases in the complexity of existing diseases.

\begin{table}[htbp]\caption{Disease-income complexity index and the complexity of diseases}
\label{tab:t7}
\begin{center}
{
\begin{tabular}{l*{2}{c}}
\toprule
                    &\multicolumn{1}{c}{(1)}&\multicolumn{1}{c}{(2)}\\
                    &\multicolumn{1}{c}{\thead{DCI\\Within Estimation}}&\multicolumn{1}{c}{\thead{DCI\\Between Estimation}}\\
\midrule
DICI                 &    0.479***   &   2.041***      \\
                    &    (0.149)   &    (0.045)    \\
\hline
Observations        &       5,211   &       5,211   \\
Diseases            &          193     & 193           \\
R-square                &     0.90     &    0.88      \\
\bottomrule
\end{tabular}
}
\end{center}
\footnotesize Notes: {\textrm DICI}: Disease-Income Complexity Index; {\textrm DCI}: Disease Complexity Index. Time dummies are included in the within regression. Standard errors are in parentheses. * p<0.10, ** p<0.05, *** p<0.01
\label{tab:7}
\end{table}

\section{Conclusions}
\label{sec: concl}

Our analysis illustrates that a country's level of development determines the structure of its disease space. Following the economic complexity methodology, we developed the {\textrm HCI}, which quantifies the network representation of the relatedness and proximity of diseases. In a dynamic panel data setting, we showed that there is a robust positive effect of a country's economic development, measured by GDP per capita, on its level of health complexity, i.e., on the `structural' composition of its pool of diseases. The evidence presented here suggests that the economic development of nations conditions the disease space. Specifically, more complex diseases tend to have relatively more prevalent cases in populations with a higher income per capita. Explicitly, it seems that when an economy accelerates, the impact on health complexity is positive. 

In addition, we build the {\textrm DICI}, which links a disease to the average level of income per capita of the countries in which the disease has prevalent cases and illustrate how disease complexity is related to economic development. Specifically, we show how changes in GDP per capita are associated with more complex diseases. The temporal variation of the above indices is important from a policy perspective. Using the {\textrm HCI} and {\textrm DCI}, it is possible to design policies aimed at improving the recognition, visibility, and traceability of complex diseases across the globe and through time (e.g., by developing a classification system for all health information systems). These indices can also be used as tools for the development of national plans for complex diseases and the establishment of knowledge networks on complex diseases, so as to improve their diagnosis, treatment, and cure. Furthermore, the {\textrm DICI} could be used to design a health expenditure reallocation policy promoting health activities and services associated with the prevention of complex diseases. 

This study employs the economic complexity methodology to compute two new metrics that quantify the disease space of countries. These can be valuable tools for estimating the effect of economic development on the health status of nations. The topic of economic complexity is a rather new one, and its use in economics is rather limited so far. By focusing on the topic of disease complexity, our contribution lies in bridging the health economics literature with the literature that highlights economic complexity as a powerful paradigm in understanding key issues in economics, geography, innovation studies, and other social sciences. 

\bibliographystyle{sg-bibstyle}
\bibliography{GLG-ComplexHealth}

\begin{thebibliography}{111}
\bibitem[{Zahm and Blair(2003)}]{zahm2003occupational}
Zahm, S.~H.; Blair, A. (2003).
\newblock Occupational cancer among women: where have we been and where are we
  going?
\newblock \emph{American journal of industrial medicine} \textbf{44(6)},
  565--575.
\end{thebibliography}

\end{document}